\documentclass[twocolumn,showpacs,superscriptaddress,longbibliography,nofootinbib,pra]{revtex4-1}
\usepackage[utf8]{inputenc}
\usepackage{calrsfs}
\usepackage{graphicx}
\usepackage{textcomp}
\usepackage{amsmath,amssymb}
\usepackage{color}
\usepackage{soul}
\usepackage{mathtools}
\usepackage{multirow}
\usepackage{subcaption}
\usepackage[colorlinks=true, urlcolor=blue, citecolor=blue, linkcolor=blue]{hyperref}
\setstcolor{red}
\begin{document}
\title{ Characteristic, dynamic, and near saturation regions of Out-of-time-order correlation in Floquet Ising models}
\author{Rohit Kumar Shukla}
\email[]{rohitkrshukla.rs.phy17@itbhu.ac.in}
\affiliation{Department of Physics, Indian Institute of Technology (Banaras Hindu University), Varanasi - 221005, India}
\author{Sunil Kumar Mishra}
\email[]{sunilkm.app@iitbhu.ac.in}
\affiliation{Department of Physics, Indian Institute of Technology (Banaras Hindu University), Varanasi - 221005, India}
\date{\today}
\begin{abstract}
 We study characteristic, dynamic, and saturation regimes of the out-of-time-order correlation (OTOC) in the constant field Floquet system with and without longitudinal field. In the calculation of OTOC, we take local spins in longitudinal and transverse directions as observables which are local and non-local in terms of Jordan-Wigner fermions, respectively. We use the exact analytical solution of OTOC for the integrable model (without longitudinal field term) with transverse direction spins as observables and numerical solutions for other integrable and nonintegrable cases.  OTOCs generated in both cases depart from unity at a kick equal to the separation between the observables when the local spins in the transverse direction and one additional kick is required when the local spins in the longitudinal direction. The number of kicks required to depart from unity depends on the separation between the observables and is independent of the Floquet period and system size. In the dynamic region, OTOCs show power-law growth in both models, the integrable (without longitudinal field) as well as the nonintegrable (with longitudinal field).  The exponent of the power-law increases with increasing separation between the observables. Near the saturation region, OTOCs grow linearly with a very small rate.  
\end{abstract}
\maketitle
\section{Introduction}
\label{Introduction}
 Larkin and Ovchinnikov first introduced the concept of out-of-time-order correlation (OTOC) for defining approaches from quasi-classical  to quantum systems \cite{larkin1969quasiclassical}.  In recent years OTOCs have gotten the attention in various fields \cite{singh2022scrambling,kitaev2014hidden, shenker2015stringy,gutzwiller2013,haake1991,hosur2016chaos,Rozenbaum2017,Garc2018,shukla2021,garcia2018,rozenbaum2020} such as quantum chaos and information propagation in quantum many-body systems \cite{maldacena2016bound, stanford2016many,aleiner2016microscopic,roberts2016lieb, roberts2015localized,bilitewski2018temperature,das2018light}, quantum entanglement and quantum-information delocalization  \cite{hosur2016chaos,huang2017out,wei2018exploring,lin2018out,abeling2018analysis,daug2019detection,grozdanov2018black}, static and dynamical phase transitions  \cite{heyl2018detecting,chen2020detecting, shukla2021}. Several proposals for experimental measurement of OTOC are proposed  \cite{yao2016interferometric,swingle2016measuring,zhu2016measurement,campisi2017thermodynamics,halpern2017jarzynski} using cold atoms or cavity and circuit quantum electrodynamics (QED) or trapped-ion simulations. Experimental realisations have been made using nuclear spins of molecules \cite{wei2018exploring,chen2020detecting,li2017measuring}, trapped ions \cite{landsman2019verified,joshi2020quantum}, and ultra-cold gases \cite{meier2019exploring}.
  Chaotic characteristics of OTOC are manifested if a small disturbance in the input of the system provides exponential deviation to the output of the system which is known as butterfly effect \cite{bilitewski2018temperature,gu2016}. 
 
 Classical Hamiltonian systems which have the highest amount of randomness and chaos are converted into the quantum domain for seeing the behaviour of quantum chaos \cite{gutzwiller2013,haake1991}.
  OTOC finds a role in characterizing the quantum chaos in these systems. There exist a characteristic form of growth of the OTOC that can distinguish different classes of information scrambling. In a chaotic case, OTOC grows very fast, which is often described by an exponential
behavior with a Lyapunov exponent. If the chaos is absent, the growth of OTOC can be much slower or even absent. In disordered systems, OTOC
distinguishes many-body localization \cite{Oganesyan2007,altman2015universal,nandkishore2015many} from the Anderson
localization \cite{anderson1958absence}.
\par

 Growth of OTOC is  also discussed in spin systems  \cite{lin2018out,xu2020accessing,xu2019locality,kukuljan2017weak,Fortes2019,Craps2020,roy2021entanglement,yan2019similar,bao2020out,dora2017out,Riddell2019,Lee2019}. Power-law growth of OTOC is observed in dynamic region of Luttinger liquid model \cite{dora2017out}, XY model \cite{bao2020out}, integrable
quantum Ising chain \cite{lin2018out} and some
systems exhibiting many-body localization \cite{Riddell2019,Lee2019}. Similar studies have been done in the Ising model with tilted magnetic fields, perturbed XXZ model, and Heisenberg spin chain with random magnetic fields \cite{Fortes2019}. In these systems, OTOC is calculated for different types of observables. For the observables that are local, non-local or mixed in terms of the Jordon-Wigner (JW) fermions, the OTOCs grow as power-law in time \cite{lin2018out}. 
\par
The quantum systems periodically driven by external forces received considerable attention for a very long time. Examples are:  kicked-rotor model in which particle moving on a ring and field is applied in the form of kicks \cite{casati1979lecture}, Chirikov standard map \cite{chirikov1971research} and the Kapitza pendulum \cite{kapitza1951dynamic}. These systems show transition from  integrability to chaos, dynamical localization \cite{Rahav2003,Rahav2003t} and dynamical stabilization \cite{kapitza1951dynamic,landau2007mechanics}. In recent years, in the quantum domain such as time crystal \cite{zhang2017observation, Russomanno2017}, the topological system with ultra-cold atoms \cite{wintersperger2020realization, Zhu2021}, a periodically-driven quantum system such as particle moving in a modulated harmonic trap
\cite{PhysRevX.4.031027}, kicked quantum rotors \cite{PhysRevLett.80.4111, PhysRevLett.87.074102,d2013many}, Floquet spin systems with constant fields \cite{gritsev2017,lakshminarayan2005,d2014long, naik2019controlled, shukla2021,Mishra2015} and quenched fields \cite{mishra2014resonance, Rossini2010, essler2016quench,Russomanno2012,Russomanno2013} got considerable attention. Periodic perturbation can be realised in experiments to understand specific properties of matter \cite{ovadyahu2012suppression,iwai2003ultrafast,kaiser2014optically, PhysRevLett.80.4111}. OTOC generated by the sum of quadratic and composite observables in terms of  Majorana fermions is studied in integrable and nonintegrable kicked quantum Ising system \cite{kukuljan2017weak} shows linear growth with time and starts to saturate at $t \simeq N/2$, where, $N$ is the system size. OTOCs using local and nonlocal observables for Floquet XY and synchronized Floquet XY
models are also studied recently  \cite{zamani2022out}. In our previous study \cite{shukla2021}, we could get the phase structure using time-averaged longitudinal magnetization OTOC (LMOTOC) but transverse magnetization OTOC (TMOTOC) failed to give us the phase diagram. While thoroughly understanding the comparison between the initial and the time-averaged behavior of integrable TMOTOC and LMOTOC, we found the different characteristic times.
In this paper, we carry out a comprehensive study of the entire region of OTOC in the integrable as well as nonintegrable Floquet spin models, not just the initial time or averaged behaviour.
We will analyse whether the integrability breaking term changes the growth of OTOC.  We extract the differences and similarities of TMOTOC and LMOTOC for integrable and nonintegrable models.
\par

This manuscript is structured as follows: In the section (\ref{model}), we will discuss the Floquet transverse Ising models. Subsequently, in the section (\ref{OTOC}), we will define transverse magnetisation OTOC (TMOTOC) and longitudinal magnetisation  OTOC (LMOTOC).  Later, we discuss results in section \ref{result}, while comparing the calculations of integrable and nonintegrable Floquet transverse Ising models in both TMOTOC and LMOTOC. Finally, we conclude the results in section \ref{conclusion}.

\maketitle
\section{Model}
\label{model}
Consider a periodically driven interacting transverse Ising Floquet system. The Hamiltonian of the system is given as 
\begin{eqnarray}
\hat H(t)=J_x\hat H_{xx}+h_{z}\sum_{n=-\infty}^{\infty}\delta\Big(n-\frac{t}{\tau}\Big) \hat H_z,
\label{int_hamiltonian}
\end{eqnarray}
where, $J_x$ is the nearest-neighbour exchange coupling strength and $h_z$ is the external field in the transverse direction applied in the form of kicks at equal interval of time $\tau$. $\hat H_{xx}=\sum_{l=1}^{N}\hat \sigma^l_x\hat \sigma^{l+1}_x $ is the nearest-neighbor Ising interaction term and $\hat H_z=\sum_{l=1}^{N}\hat \sigma^l_z$ is the interaction of unit magnetic field with the total transverse magnetization. 
\par 
Floquet map corresponding to the Eq.(\ref{int_hamiltonian}) is 
\begin{eqnarray}
\label{U0}
 \mathcal{\hat U}_0=\exp(-i \tau J_x \hat H_{xx})  \exp( -i \tau h_{z} \hat H_{z}),
\end{eqnarray}
Since in Eq.(\ref{int_hamiltonian}) there is only transverse field is present and the Hamiltonian is exactly solvable using Jordan-Wigner (JW) transformation \cite{lakshminarayan2005,Prosen2000,Prosen2002}. Now, if we introduce a longitudinal field term $  h_x \hat H_x=h_x\sum_{l=1}^{N}\hat \sigma^l_x$, the Hamiltonian can be written as  
\begin{eqnarray}
\hat H(t)=J_{x}\hat H_{xx}+h_{x}\hat H_x+h_{z}\sum_{n=-\infty}^{\infty}\delta\Big(n-\frac{t}{\tau}\Big) \hat H_z.
\label{H0}
\end{eqnarray}
However, the model could not be transformed into the free fermions using JW transformation because the longitudinal field term when transformed into JW fermions gives an interacting fermionic term \cite{Prosen2000,Prosen2002}. 
The Floquet map corresponding to this model is
\begin{eqnarray}
\label{Ux}
 \mathcal{\hat U}_x=\exp\big[-i \tau (J_x \hat H_{xx}+h_{x}\hat H_{x})\big]  \exp( -i \tau h_{z} \hat H_{z}).
\end{eqnarray}
Henceforth in the manuscript, we mean integrable transverse Ising Floquet model as $\mathcal{\hat U}_0$ and nonintegrable transverse Ising Floquet model as $\mathcal{\hat U}_x$.  

\section{ TMOTOC and LMOTOC}
  \label{OTOC}
Let us consider a pair of observables $\hat W^l$ and $\hat V^m$ at $l$th and $m$th sites, respectively. OTOC of these observables is defined as
\begin{eqnarray}
C^{l,m}(n)&=&-\frac{1}{2}\langle[\hat W^l(n), \hat V^m(0)]^\dagger[\hat W^l(n), \hat V^m(0)] \rangle.
\end{eqnarray}
Observables, $\hat W^l$ and $\hat{V}^m$ are separated by distance $\Delta l=\vert l-m \vert$. Initially at $n=0$, both the observables commute to each other {\it{i.e.}} $[\hat W^l(0), \hat V^m(0)]=0$. As time increases, higher order terms of the time evolution of $\hat W^l(0)$ given by Baker-Campbell-Hausdorff formula  fail to commute with $\hat V^m$, resulting noncommutative  $\hat W^l(n)$ and $\hat{V}^m$. By examining noncommutativity of $\hat V^m$ at different positions, one can quantify upto some degree how $\hat W^l(n)$ spread over the space. Here $\hat W^l(n)$ is  $ (\mathcal{\hat U}_{x/0}^{\dagger})^n\hat W^l(0)  (\mathcal{\hat U}_{x/0})^n$.  If $\hat W^l$ and $\hat V^m$ are Hermitian and unitary, the OTOC simplifies in the form
\begin{eqnarray}
\label{gene_OTOC}
C^{l,m}(n)=1-\Re[F^{l,m}(n)],
\end{eqnarray}
where, $F^{l,m}(n)=\langle \hat W^l(n) \hat V^m(0) \hat W^l(n) \hat V^m(0) \rangle$ and $\langle\cdot \rangle$, denotes the quantum mechanical  average  over  the  initial state.
\par
OTOC is calculated with either trace over a maximally mixed state or a thermal ensemble. Trace can be replaced by employing Haar random states of $2^N$ dimensions to evaluate expectation values, that is 
$\mbox{Tr}(\hat W^l(n) \hat V^m(0) \hat W^l(n) \hat V^m(0))/2^N \approx \left \langle \Psi_R\big\vert\hat W^l(n) \hat V^m(0) \hat W^l(n) \hat V^m(0)\big\vert\Psi_R\right \rangle$
where $|\Psi_R\rangle$ is a random state.  We replaced random state by two fully polarized special initial states according to the observables and find that there is no remarkable differences in the characteristic, dynamic, and saturation regions of  OTOC. We observe only one difference in the saturation region, there is comparatively small oscillations when consider random state. Detailed discussion is mentioned in the Appendix~\ref{appendix1}. Moreover, the special initial states may help to get exact analytical formula at least for integrable OTOC cases with transverse direction spins as
observables.
\par
In this manuscript we consider $\hat W^l$ and $\hat V^m$ as local Pauli operators either in the longitudinal direction $\hat \sigma^{l,m}_x$ or in the transverse direction $\hat \sigma^{l,m}_z$. For the Pauli operators in transverse direction as local observables, the OTOC, in this paper, is defined as transverse magnetization OTOC (TMOTOC) and given as: 
\begin{eqnarray}
\label{F_z}
C_z^{l,m}(n) &=&1- \Re[F_z^{l,m}(n)],
\end{eqnarray}
where, $F_z^{l,m}(n)=\langle \phi_0|\hat \sigma^l_z(n)\hat \sigma^m_z\hat \sigma^l_z(n)\hat \sigma^m_z|\phi_0\rangle$. In the fermionic representation, $\hat \sigma_z^l$ can be written as $\hat \sigma_z^{l}=-(\prod_{j<l}A^jB^j)A^l$, where, $A^l$ and $B^l$ are defined by fermionic creation ($c^{l\dagger})$ and annihilation operator ($c^l$) as, $A^l=c^{l\dagger}+c^l$  and $B^l=c^{l \dagger}-c^l$ \cite{sachdev2011}. Since $\hat \sigma_z^{l}$ contain string  operator, hence it is known as non-local operator in term of Jordan-Wigner fermion \cite{sachdev2011,lin2018out}.

For the calculation purpose we take initial state as $|\phi_0\rangle=| \uparrow  \uparrow  \uparrow \cdots  \uparrow \rangle$, 
where $ \left| \uparrow \right\rangle$ is the eigenstate of $\hat \sigma_z$ with eigenvalue $+1$.
If the observables are taken as Pauli operators in the longitudinal direction of the Ising axis ({\it i. e.,} z-axis) then OTOC will be referred to as longitudinal magnetization OTOC (LMOTOC). The LMOTOC is given as:
\begin{eqnarray}
\label{F_x}
C_x^{l,m}(n)&=&1- \Re[F_x^{l,m}(n)],
\end{eqnarray}
where,
$F_x^{l,m}(n)=\langle
 \psi_0|\hat \sigma_x^l(n)\hat \sigma_x^m\hat \sigma^l_x(n)\hat \sigma^m_x|\psi_0\rangle$. In the fermionic representation, $\hat \sigma^{l/m}_x$ can be written as $\hat \sigma_x^{l/m}=A^{l/m}B^{l/m}$. In fermionic representation $\hat \sigma_x^{l/m}$ is known as local observable \cite{sachdev2011,lin2018out}.
 In this case the initial state will be taken as  $|\psi_{0} \rangle=|\rightarrow \rightarrow \rightarrow \cdots \rightarrow \rangle$, where, $ \left| \rightarrow \right\rangle$ is the eigenstate of $\hat \sigma_x$ with eigenvalue $+1$.  
\par
Analytical solution of the TMOTOC for the initial state $|\phi_0\rangle=| \uparrow  \uparrow  \uparrow \cdots  \uparrow \rangle$ and Floque  map defined by Eq.~\ref{U0} with $J_x=1$ and $h_z=1$
is derived in the Ref.~\cite{shukla2021} as
\begin{eqnarray}
\label{OTOCz_gene}
 F_z^{l,m}(n) &=& 1- \Big(\frac{2}{ N}\Big)^3 \sum_{p,q,r} \Big[ e^{i(p-q)(m-l)}|\Psi_r(n)|^2 \Phi_p^{*}(n) \Phi_q(n) \nonumber \\
 &-& e^{i(-r-q)(m-l)}  \Psi_r(n)^{*} \Phi_p^{*}(n)  \Phi_q(n) \Psi_{-p}(n)  \nonumber \\
&-& e^{i(p+q)(m-l)} \Psi_{q}(n)  \Psi_{r}(n)^{*}  \Phi_{p}(n)^{*} \Phi_{-r}(n)\nonumber \\
&+& e^{i(q-r)(m-l)} \Psi_{q}(n)\Psi_r(n)^* |\Phi_p(n)|^2 \Big],  
\end{eqnarray}
where the expansion coefficients  $\Phi_q(n)$ and  $\Psi_q(n)$ are defined as 
\begin{equation}
\label{phi}
\Phi_q(n)=|\alpha_{+}(q)|^2 e^{-i n \gamma_q}+|\alpha_{-}(q)|^2 e^{i n \gamma_q},
\end{equation}
\begin{equation}
\label{psi}
\Psi_q(n)=\alpha_{+}(q) \beta_{+}(q)e^{-i n \gamma_q}+\alpha_{-}(q)\beta_{-}(q) e^{i n \gamma_q}.
\end{equation}
The phase angle $\gamma_q$ and the coefficients $\alpha_{\pm}(q)$ and $\beta_{\pm}(q)$ are given by
\begin{equation}
\label{gamma}
\cos(\gamma_q)=\cos(2 \tau)\cos(4\tau)-\cos(q)\sin(2\tau)\sin \Big(2\tau \Big),
\end{equation}
and
\begin{equation}
\label{apmq}
\alpha_{\pm}(q)^{-1}=\sqrt{1+\Big(\frac{\cos(2\tau)-cos(\gamma_q \pm 2 \tau)}{\sin(q) \sin(
2\tau)\sin(2\tau)}\Big)^2},
\end{equation}

\begin{eqnarray}
\label{bpmq}
\beta_{\pm}(q)&=& \frac{\mp\sin(\gamma_q)-\cos( 2\tau)\sin(2\tau)\big(\cos(q) +1\big)}{\sin(q)\sin(2\tau)}  \nonumber \\
&\times & \alpha_{\pm}(q)e^{-i2\tau}.
\end{eqnarray}
The allowed value of $p$, $q$ and $r$ are from $\frac{-(N-1)\pi}{N}$ to $\frac{(N-1)\pi}{N} $ differing by $\frac{2 \pi}{N} $ for even number of $N_F$ ($N_F=\sum_l c_l^\dagger c_l$, number of fermions) and $\hbar=1$. 
 We use the above exact solution to calculate TMOTOC for integrable $\mathcal{\hat U}_0$ model. However, TMOTOC for the nonintegrable $\mathcal{\hat U}_x$ model, and LMOTOC for both integrable $\mathcal{\hat U}_0$ and nonintegrable $\mathcal{\hat U}_x$ model will be calculated numerically.

\begin{figure}[hbt!]
 \includegraphics[width=.90\linewidth, height=.50\linewidth]{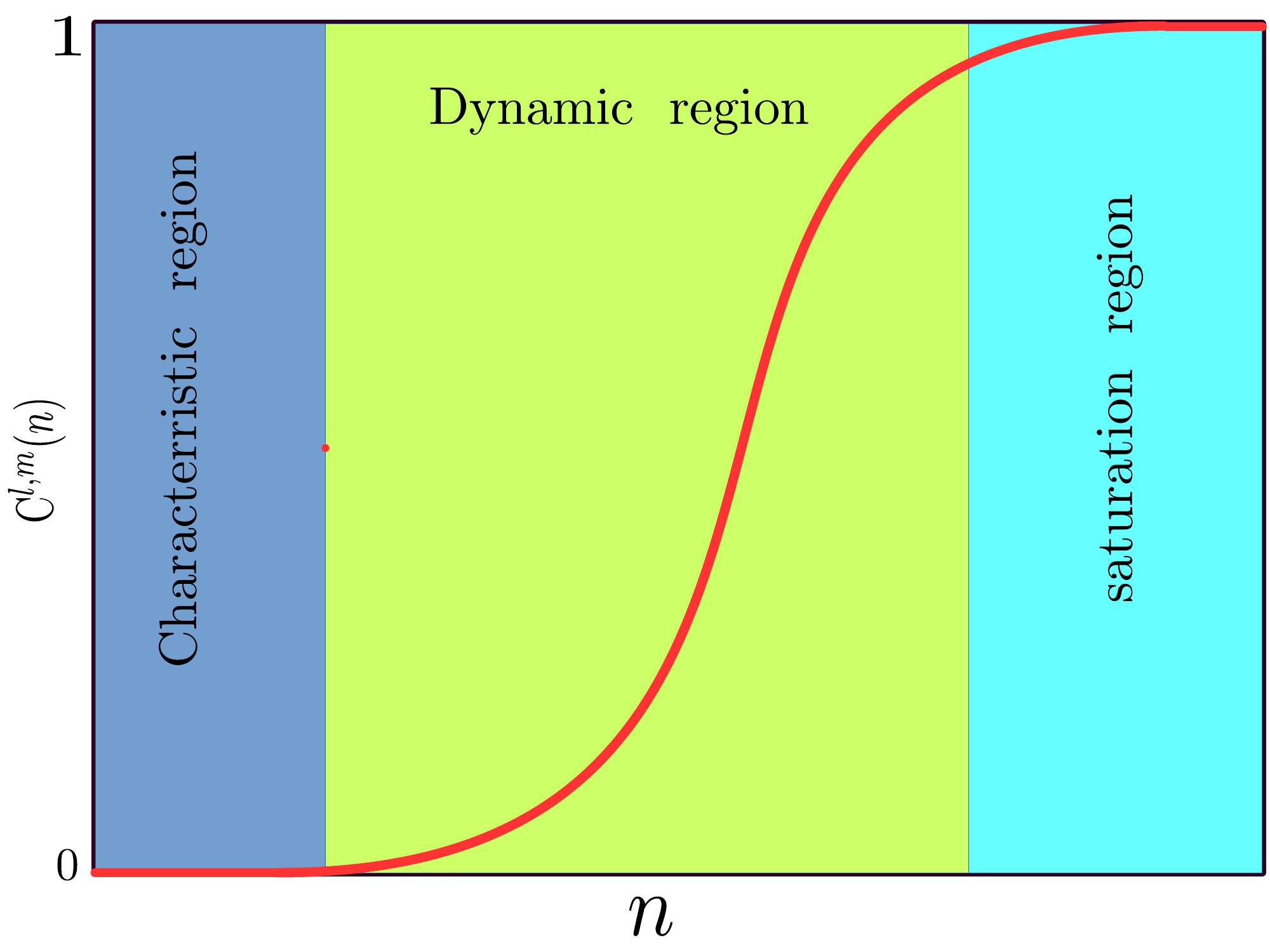}
 \caption{Schematic of the various regions of OTOC in a typical system.} 
   \label{otocfig}
  \end{figure}
\begin{figure*}[hbt!]
\begin{subfigure}[a]{.30\textwidth} 
\includegraphics[width=.99\linewidth, height=.70\linewidth]{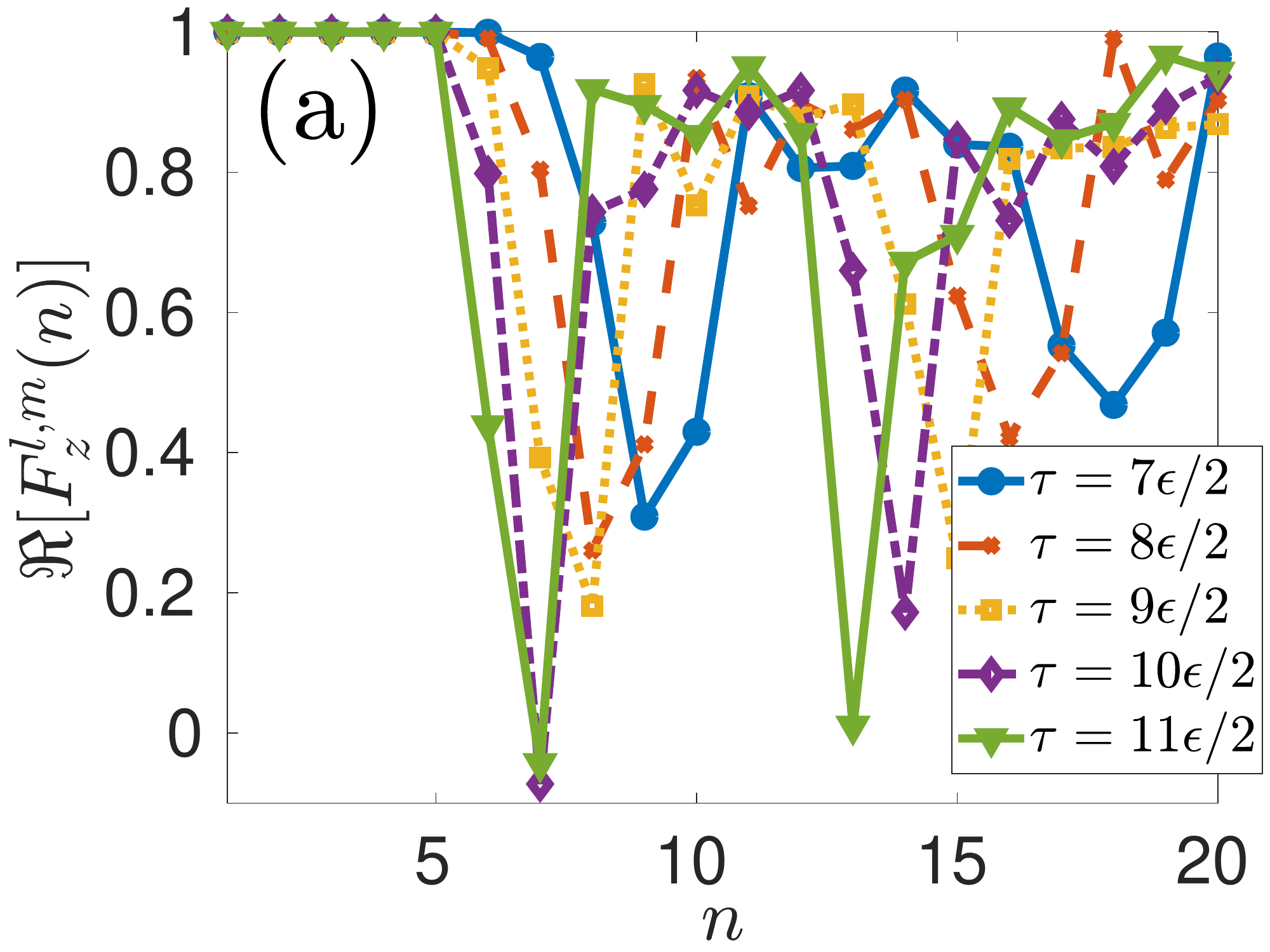}
 \end{subfigure}
 \begin{subfigure}{.30\textwidth} 
 \includegraphics[width=.99\linewidth, height=.70\linewidth]{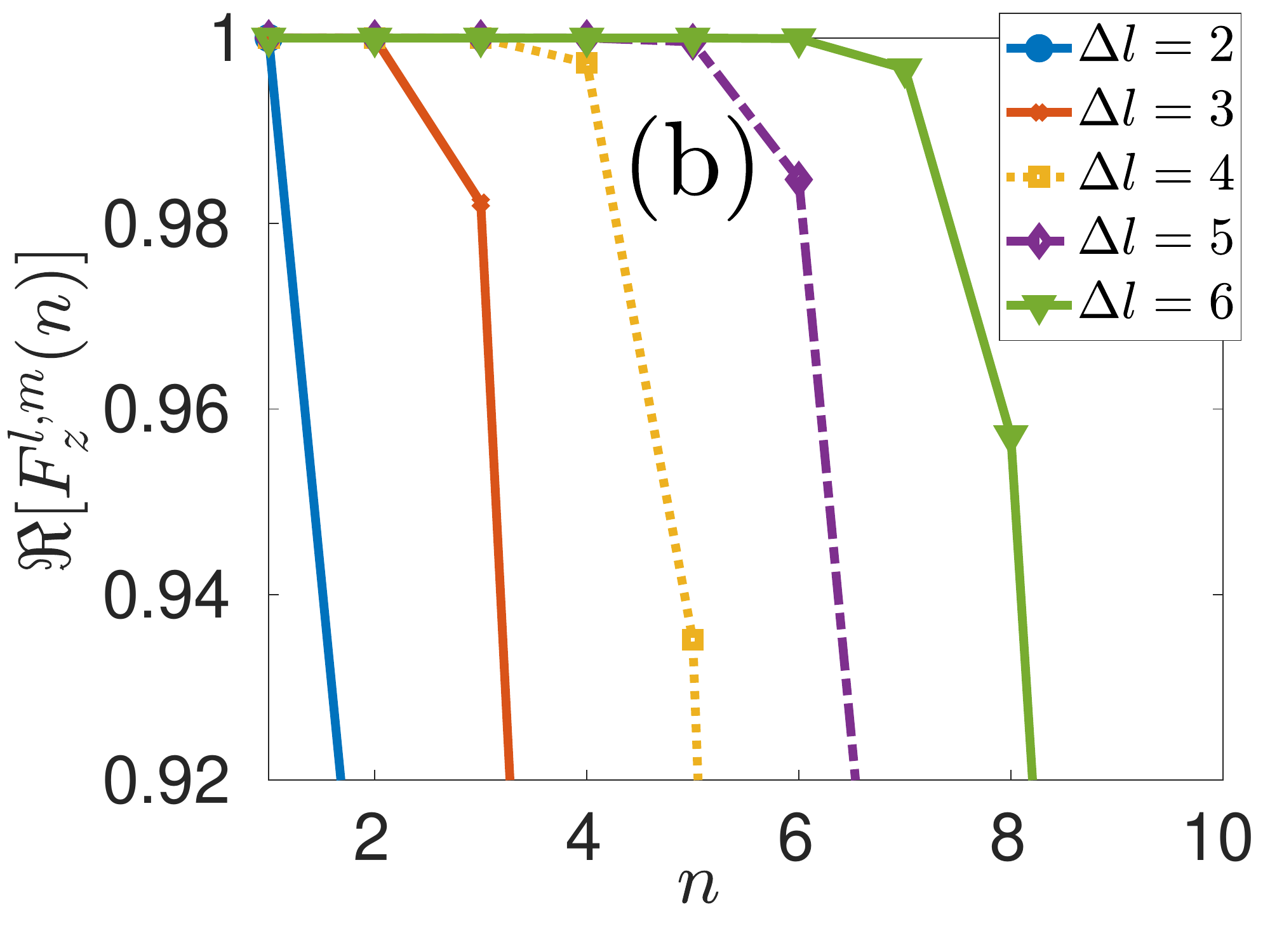}
 \end{subfigure}
 \begin{subfigure}{.30\textwidth} 
 \end{subfigure}
 \begin{subfigure}{.30\textwidth}
   \includegraphics[width=.99\linewidth, height=.70\linewidth]{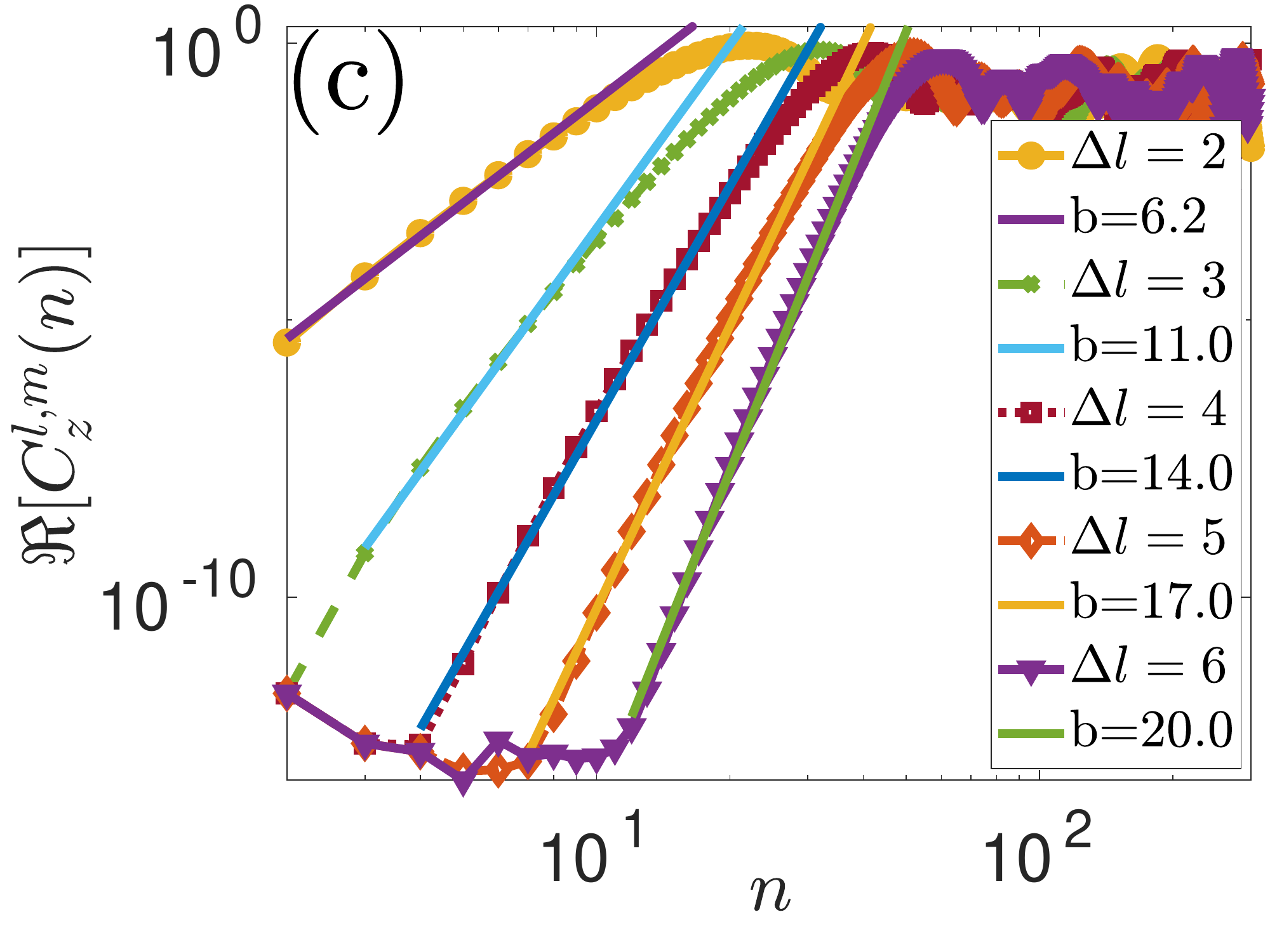}
  \end{subfigure}
  \begin{subfigure}{.30\textwidth}
   \includegraphics[width=.99\linewidth, height=.70\linewidth]{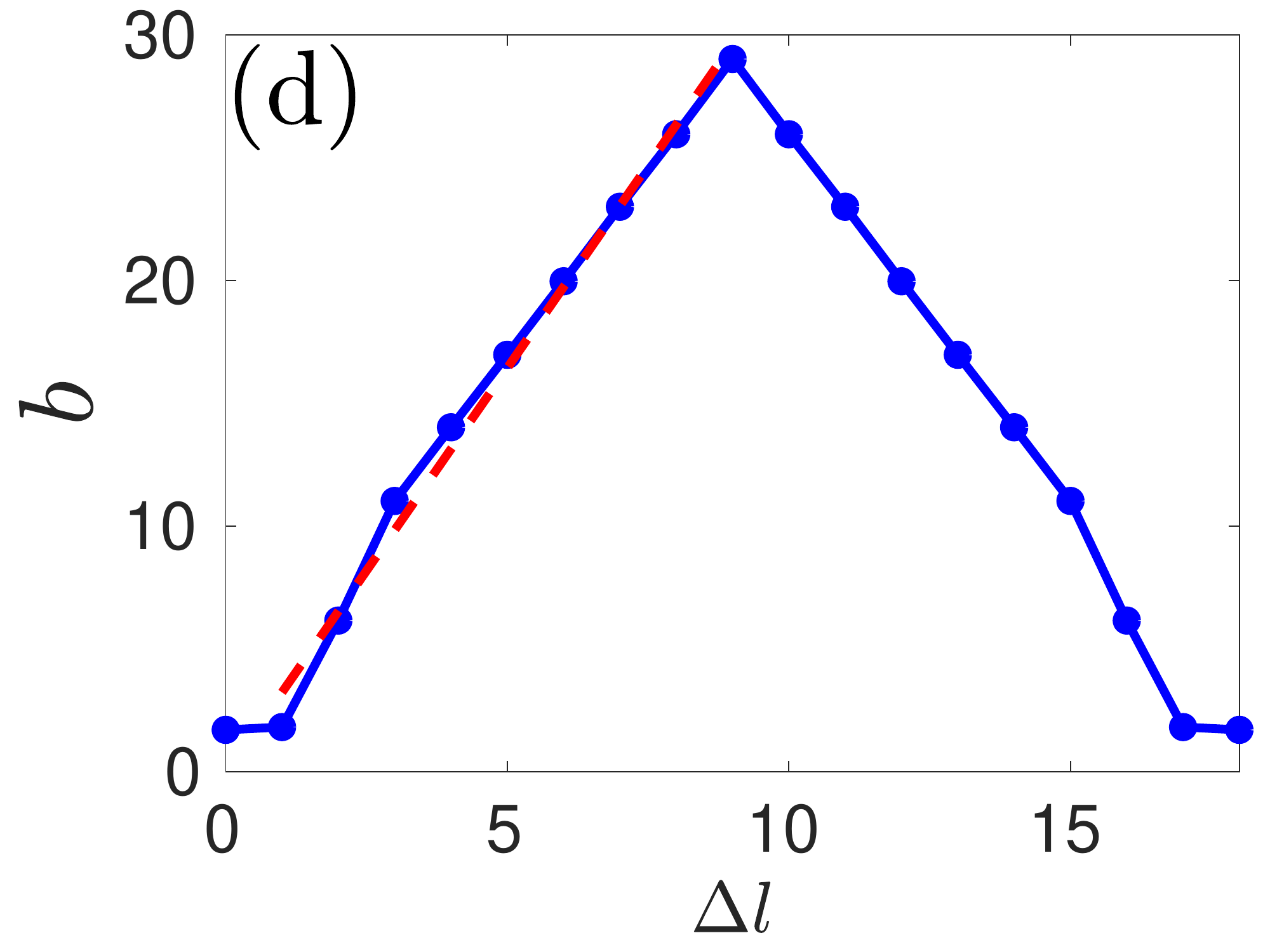}
  \end{subfigure}
  \begin{subfigure}{.30\textwidth}
   \includegraphics[width=.99\linewidth, height=.70\linewidth]{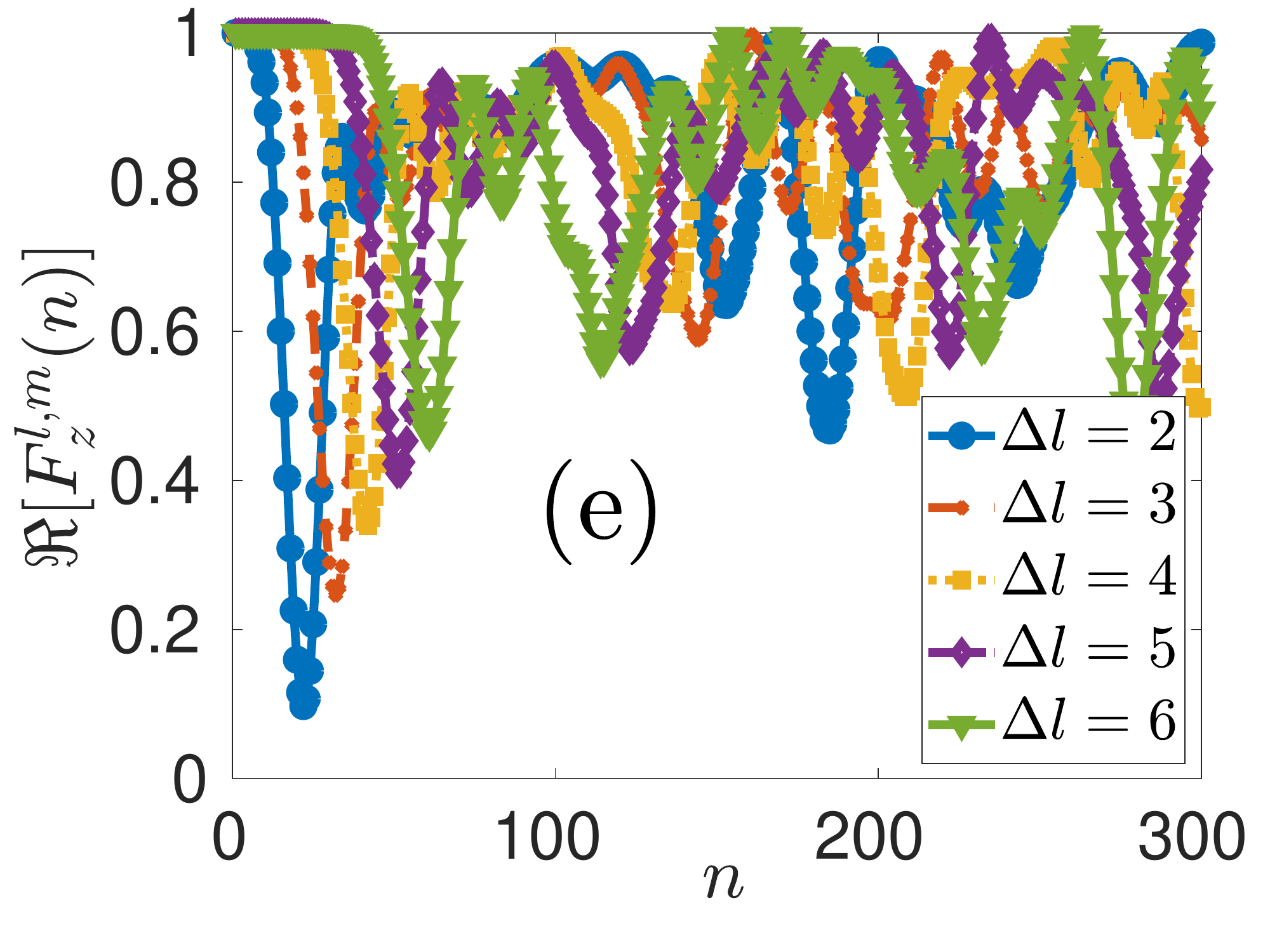}
  \end{subfigure}
 \caption{Integrable transverse Ising Floquet system  with $J_x=1$ and $h_z=1$ for $N=18$. (a) Behaviour of $TMOTOC$ with number of kicks $(n)$ by increasing value of Floquet period from 
 $\frac{7\epsilon}{2}$ to $\frac{11 \epsilon}{2}$ (right to left) differing by $\epsilon/2$ with fixed $\Delta l=6$ ($\epsilon=\frac{\pi}{28})$. (b)  $F^{l,m}_z$ with number of kicks by increasing $\Delta l$ (left to right) and fixed Floquet period  $\tau=6\epsilon/2$. (c) $C^{l,m}_z$ with number of kicks ($\log-\log$) with increasing $\Delta l$ (left to right) at constant Floquet period $\tau=\frac{\epsilon}{2}$. (d) Exponent of power-law with increasing distance between the spins. (e)  $\Re[F^{l,m}_z]$ with number of kicks at different increasing $\Delta l$ (left to right). } 
 \label{cf_TMOTOC_int}
\end{figure*}

 \begin{figure*}[hbt!]
\begin{subfigure}{.30\textwidth} 
 \includegraphics[width=.99\linewidth, height=.70\linewidth]{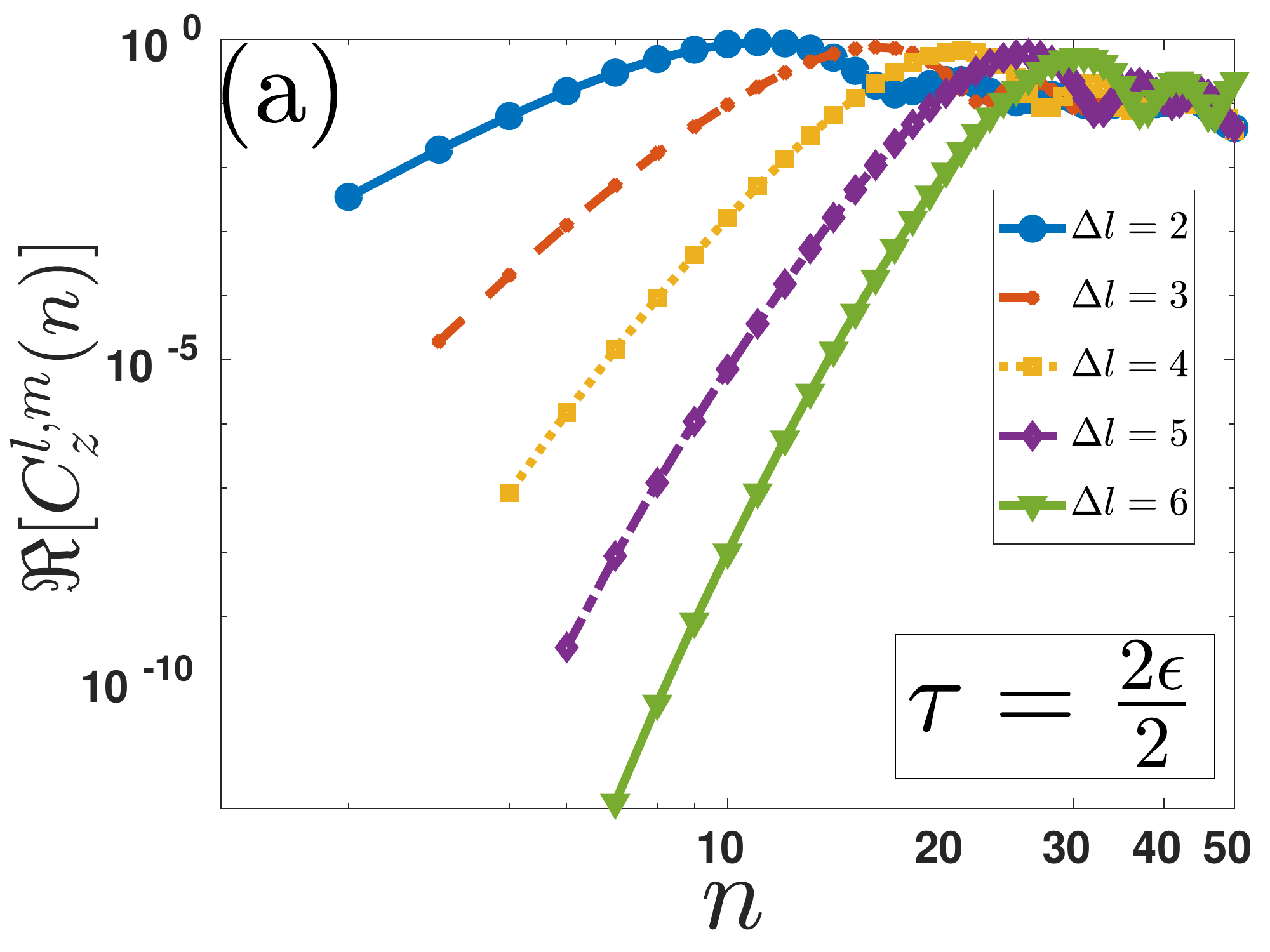}
 \end{subfigure}
 \begin{subfigure}{.30\textwidth} 
 \includegraphics[width=.99\linewidth, height=.70\linewidth]{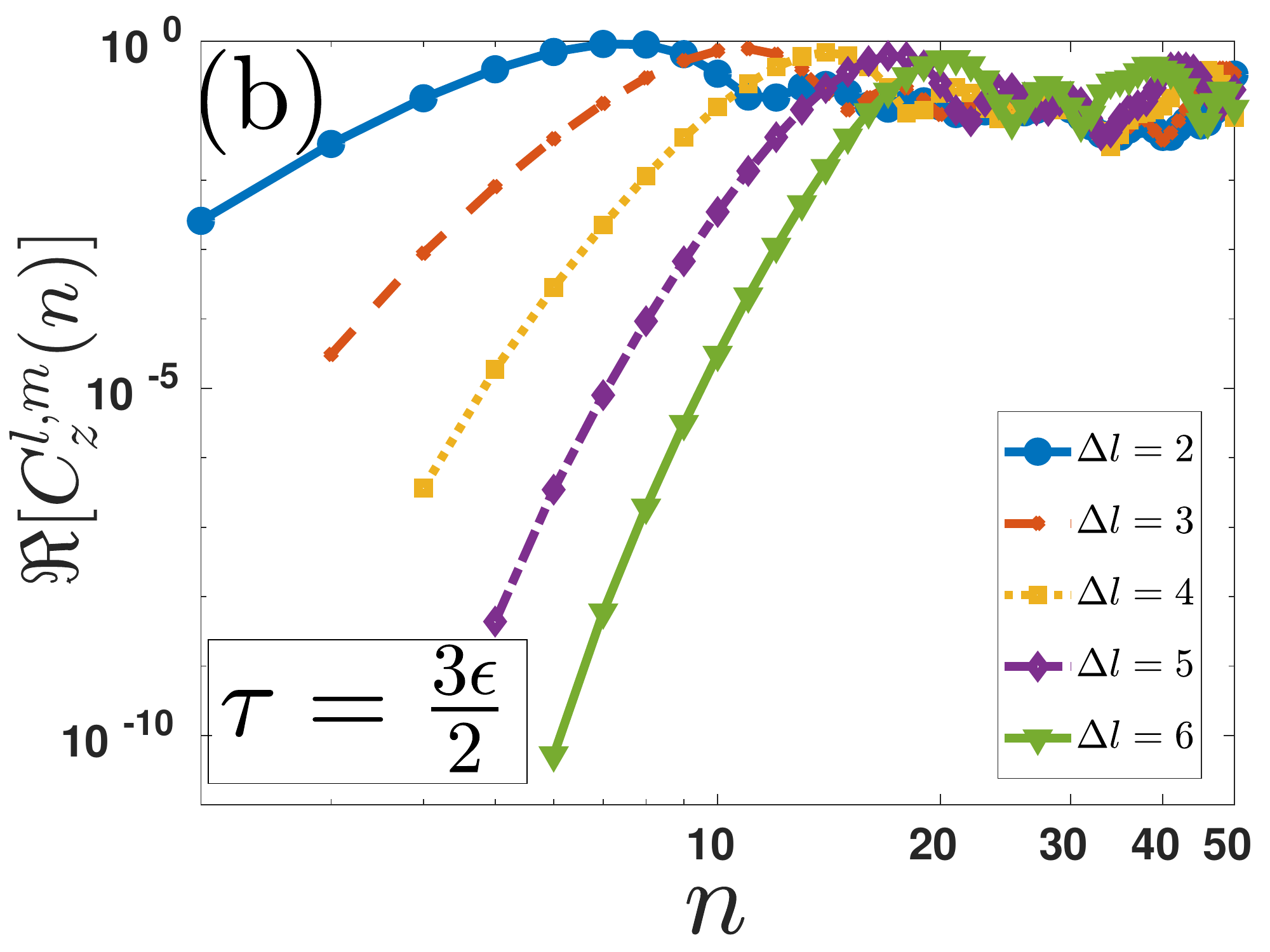}
 \end{subfigure}
 \begin{subfigure}{.30\textwidth}
   \includegraphics[width=.99\linewidth, height=.70\linewidth]{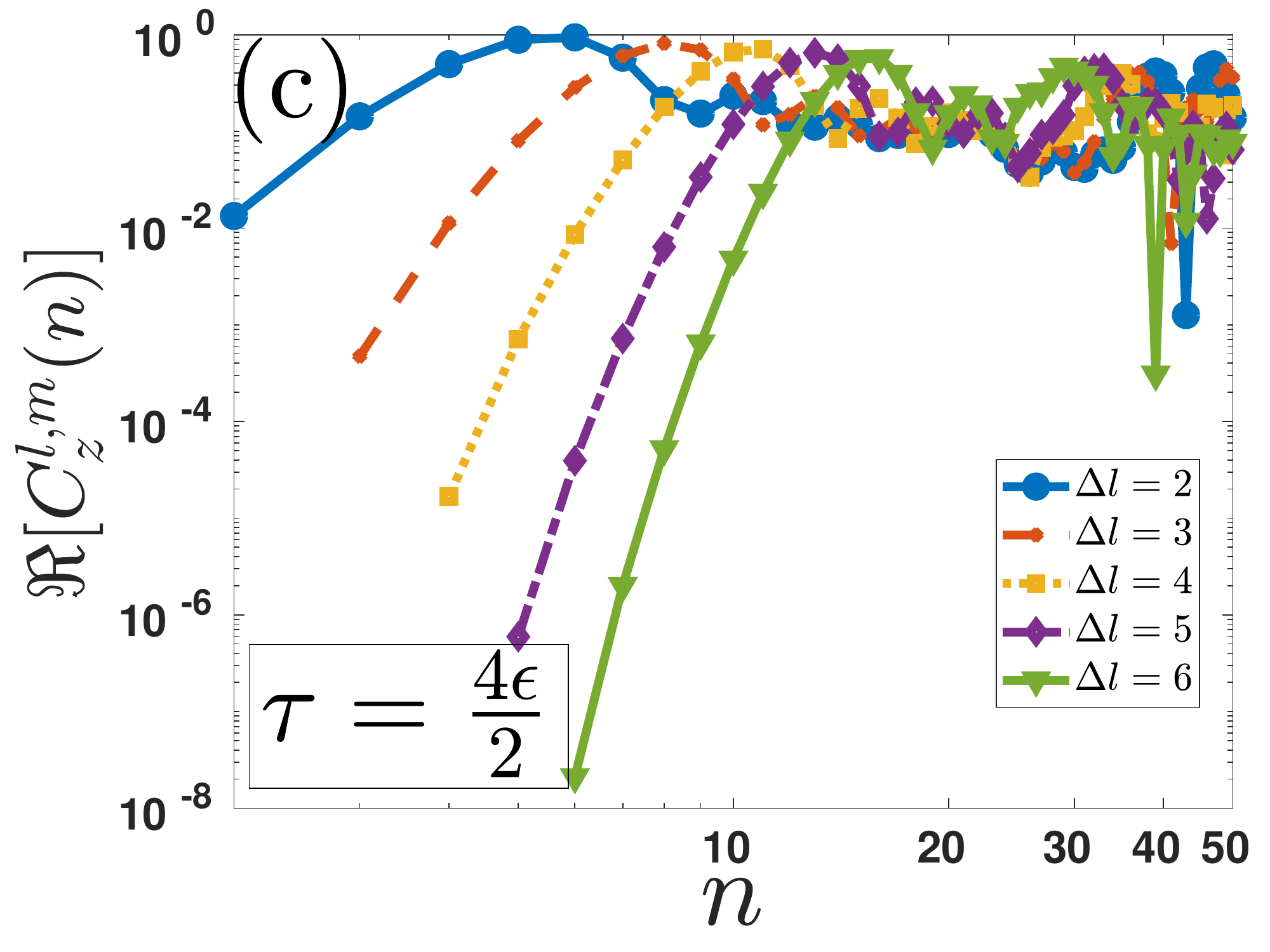}
  \end{subfigure}
  \begin{subfigure}{.30\textwidth}
   \includegraphics[width=.99\linewidth, height=.70\linewidth]{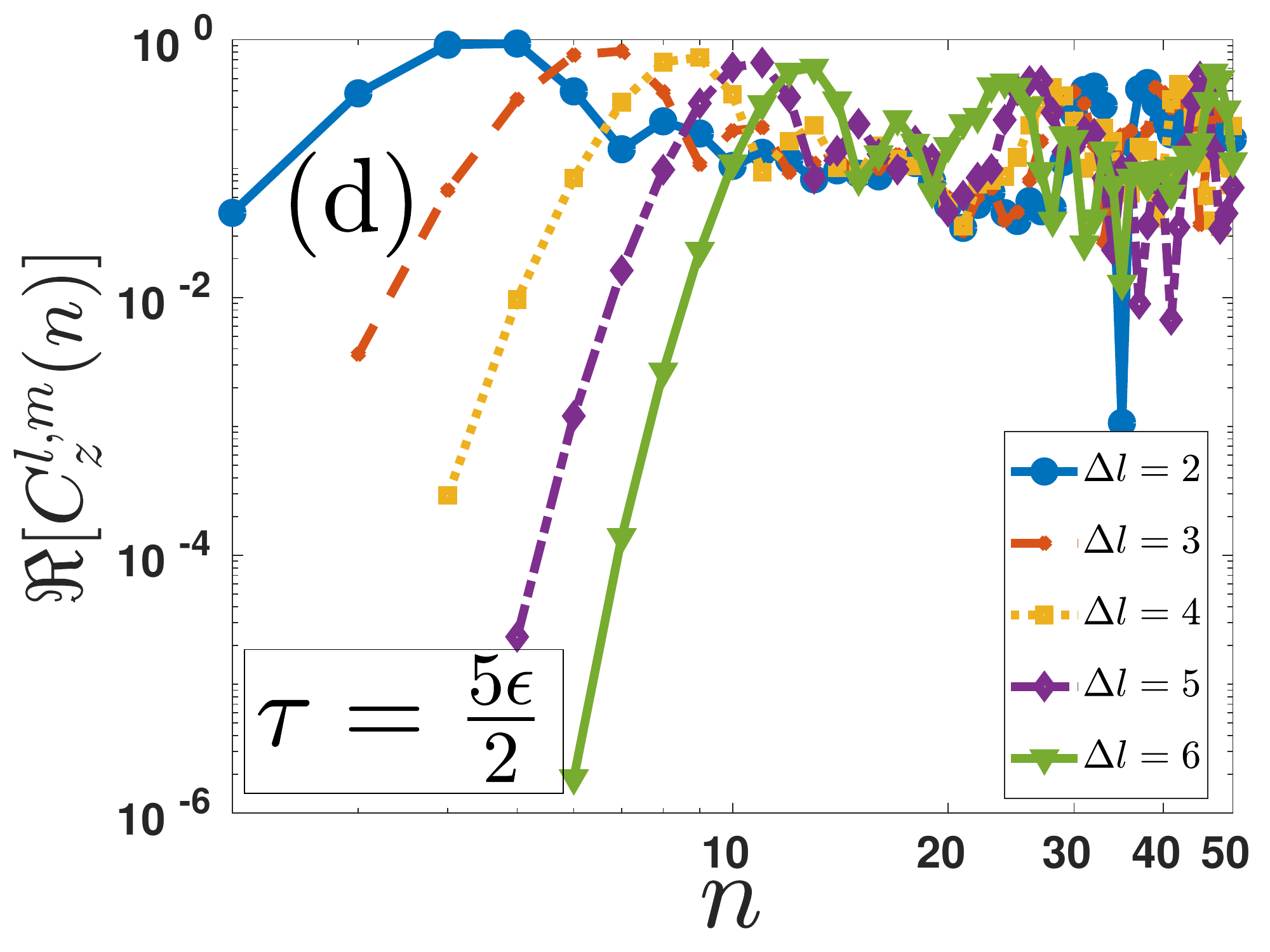}
  \end{subfigure}
  \begin{subfigure}{.30\textwidth}
   \includegraphics[width=.99\linewidth, height=.70\linewidth]{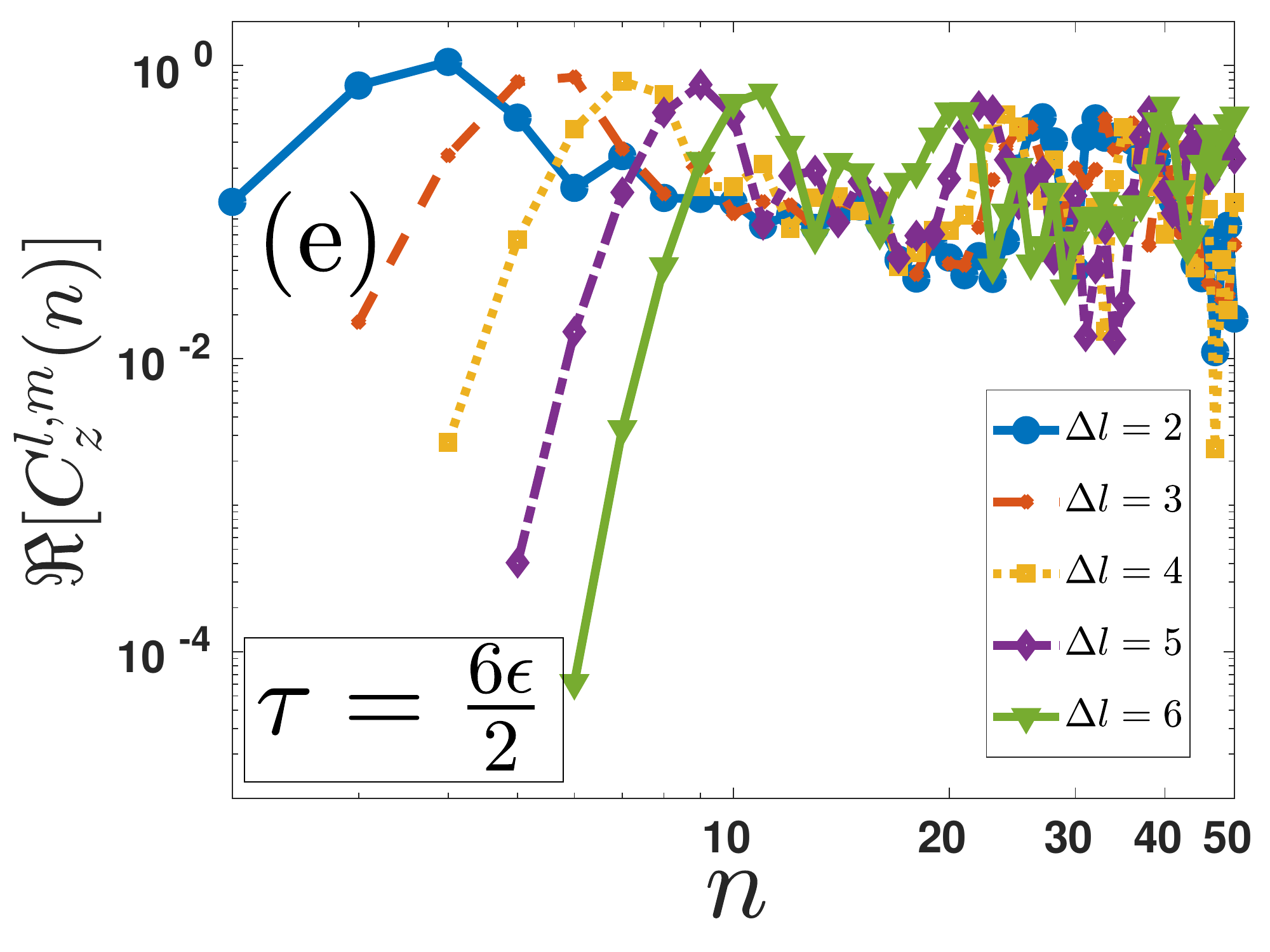}
  \end{subfigure}
  \caption{Integrable transverse Ising Floquet system  with $J_x=1$ and $h_z=1$ 
  for $N=18$.  Behaviour of $TMOTOC$ with number of kicks $(n)$  by increasing $\Delta l$ from $2$ to $6$ (left to right) at different  Floquet period (a) $\tau=\frac{2\epsilon}{2}$, (b) $\tau=\frac{3\epsilon}{2}$, (c) $\tau=\frac{4\epsilon}{2}$, (d) $\tau=\frac{5\epsilon}{2}$ and (e) $\tau=\frac{6\epsilon}{2}$ ($\epsilon=\frac{\pi}{28}$). } 
 \label{cf_TMOTOC}
\end{figure*}


\begin{figure*}[hbt!]
\begin{subfigure}{.30\textwidth} 
  \includegraphics[width=.99\linewidth, height=.70\linewidth]{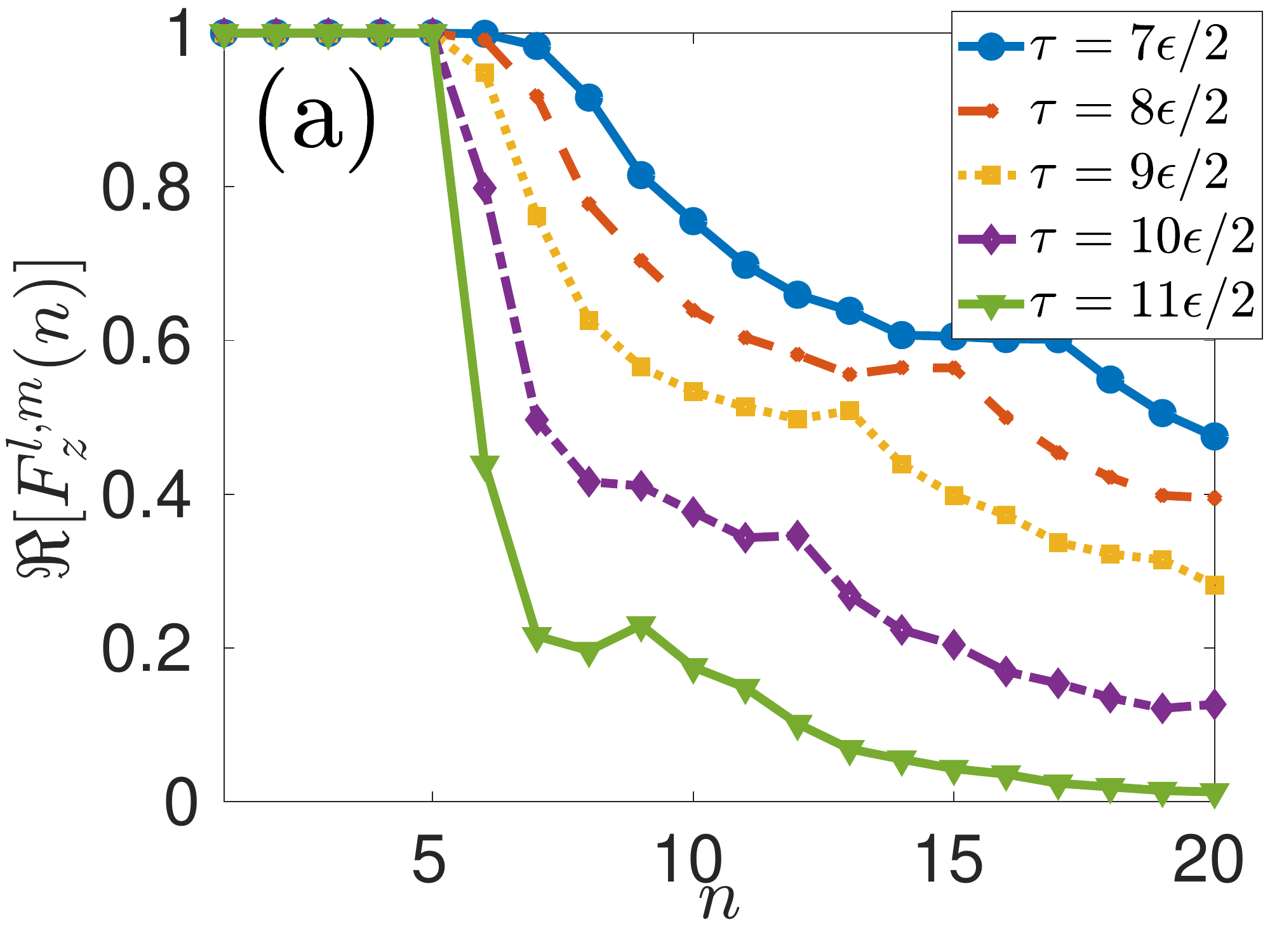}
 \end{subfigure}
 \begin{subfigure}{.30\textwidth} 
 \includegraphics[width=.99\linewidth, height=.70\linewidth]{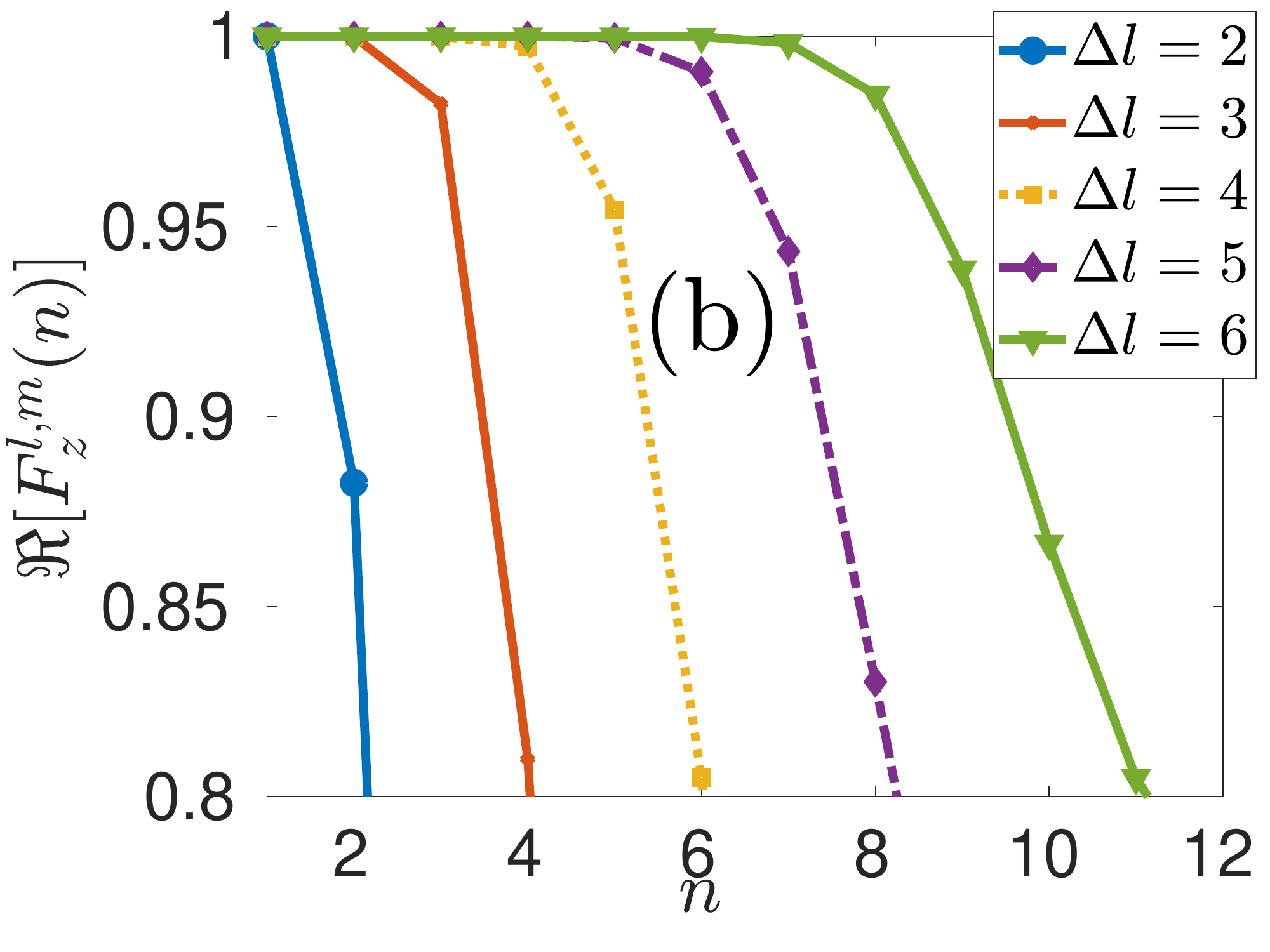}
 \end{subfigure}
 \begin{subfigure}{.30\textwidth} 
 \end{subfigure}
 \begin{subfigure}{.30\textwidth}
  \includegraphics[width=.99\linewidth, height=.70\linewidth]{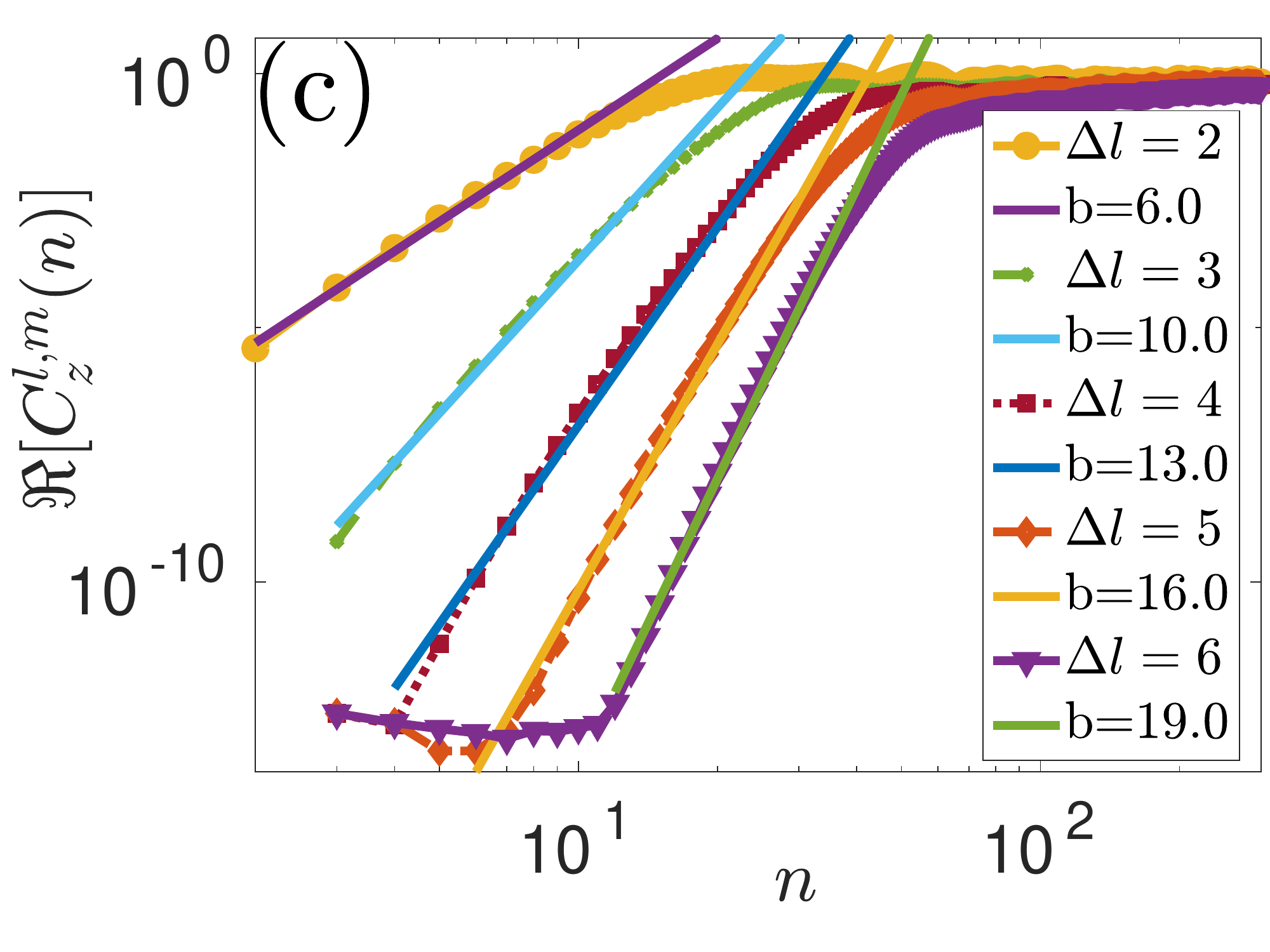}
  \end{subfigure}
  \begin{subfigure}{.30\textwidth}
  \includegraphics[width=.99\linewidth, height=.70\linewidth]{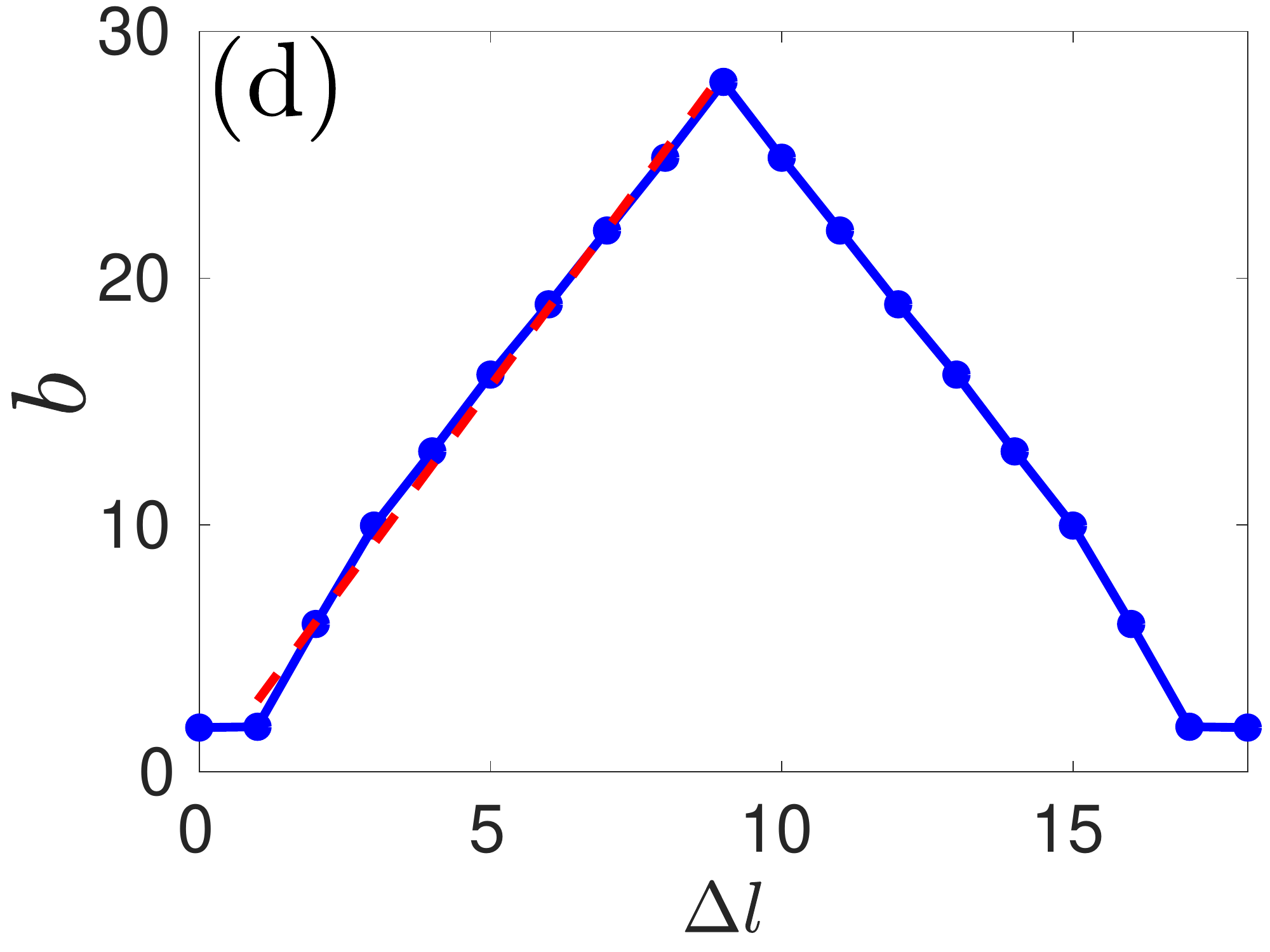}
  \end{subfigure}
  \begin{subfigure}{.30\textwidth}
  \includegraphics[width=.99\linewidth, height=.70\linewidth]{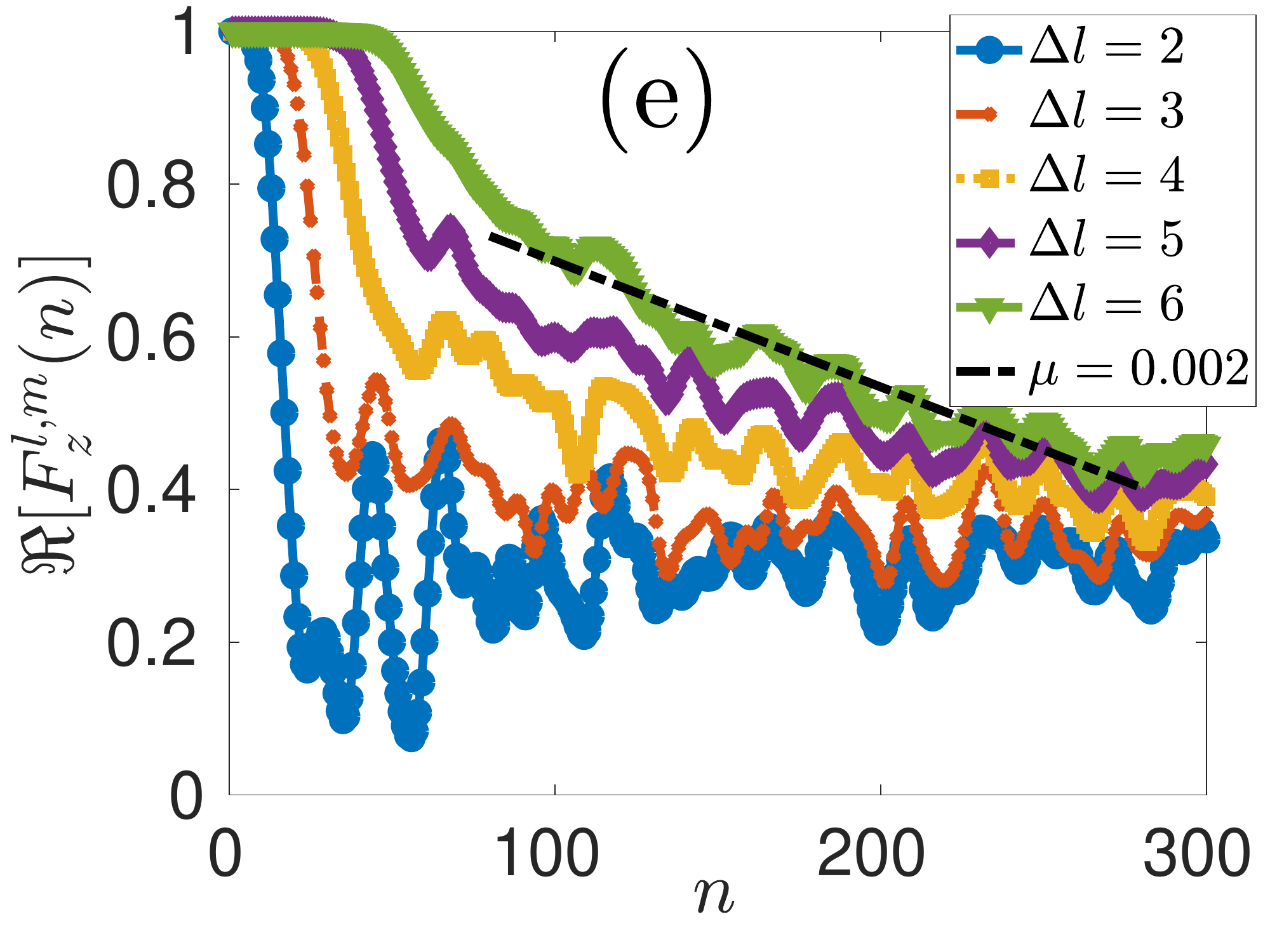}
  \end{subfigure}
 \caption{Nonintegrable closed chain transverse Ising Floquet system  with $J_x=1$, $h_z=1$, and $h_x=1$ of size $N=18$. (a) Behaviour of $TMOTOC$ with number of kicks $(n)$ by increasing value of Floquet period from  $\frac{7\epsilon}{2}$ to $\frac{11\epsilon}{2}$ (right to left) differing by  $\frac{\epsilon}{2}$  with fixed $\Delta l=6$ ($\epsilon=\frac{\pi}{28})$.  (b)  Initial region of $F^{l,m}_z$ with number of kicks with increasing $\Delta l$ (left to right) and fixed Floquet period $\tau=6\epsilon/2$. (c)   $C^{l,m}_z$ with number of kicks ($\log-\log$) with increasing $\Delta l$ (left to right) at fixed $\tau=\frac{\epsilon}{2}$.  (d) Changing of power with $\Delta l$. (e) Saturation of $F^{l,m}_z$ with number of kicks at different increasing $\Delta l$ (left to right)}. Black dotted line represent the linear decreasing of maxima of saturation amplitude.
\label{cf_TMOTOC_nint}
\end{figure*}
\begin{figure*}[hbt!]
\begin{subfigure}{.30\textwidth} 
 \includegraphics[width=.99\linewidth, height=.70\linewidth]{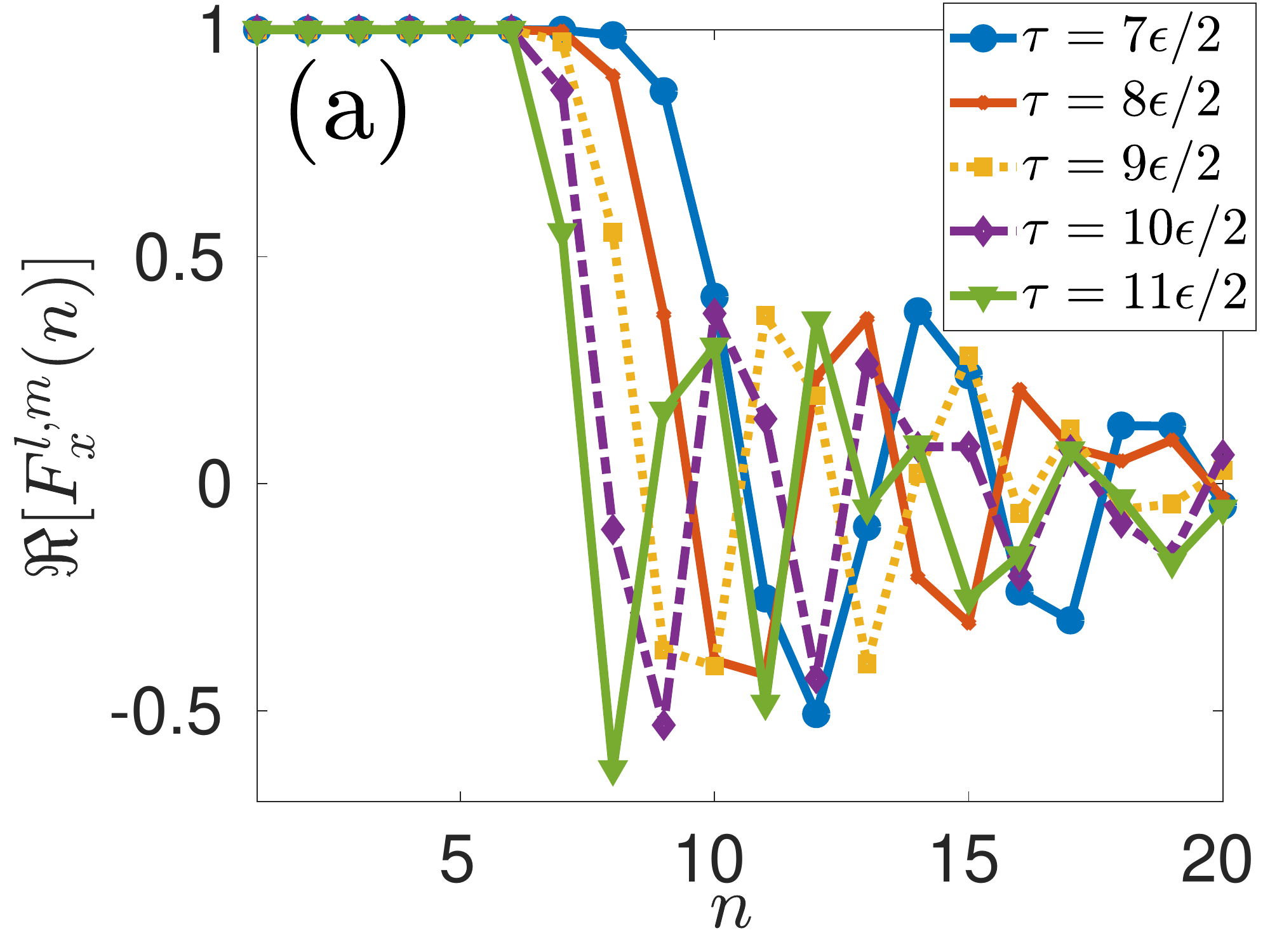}
 \end{subfigure}
 \begin{subfigure}{.30\textwidth} 
 \includegraphics[width=.99\linewidth, height=.70\linewidth]{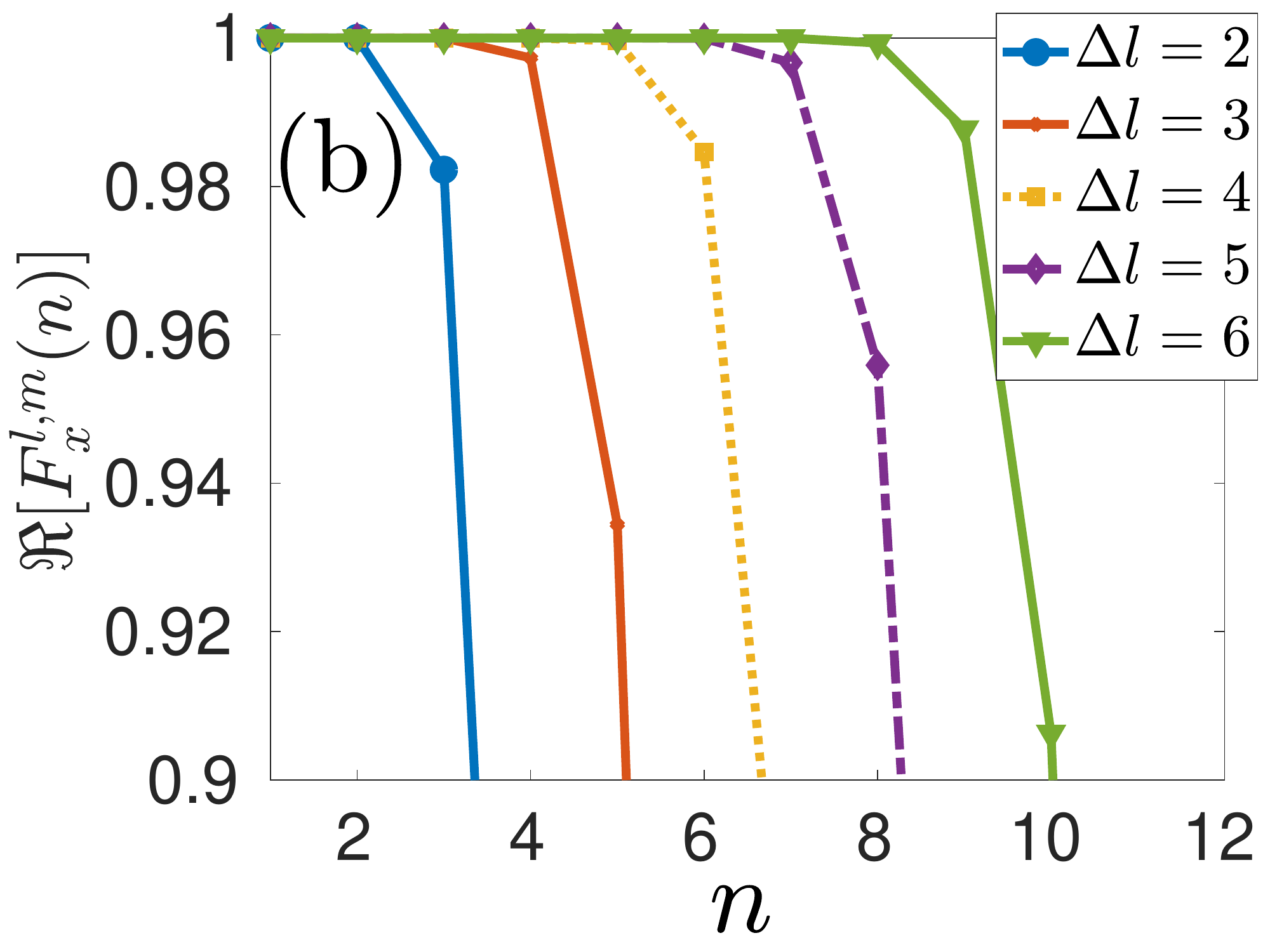}
 \end{subfigure}
 \begin{subfigure}{.30\textwidth}
  \includegraphics[width=.99\linewidth, height=.70\linewidth]{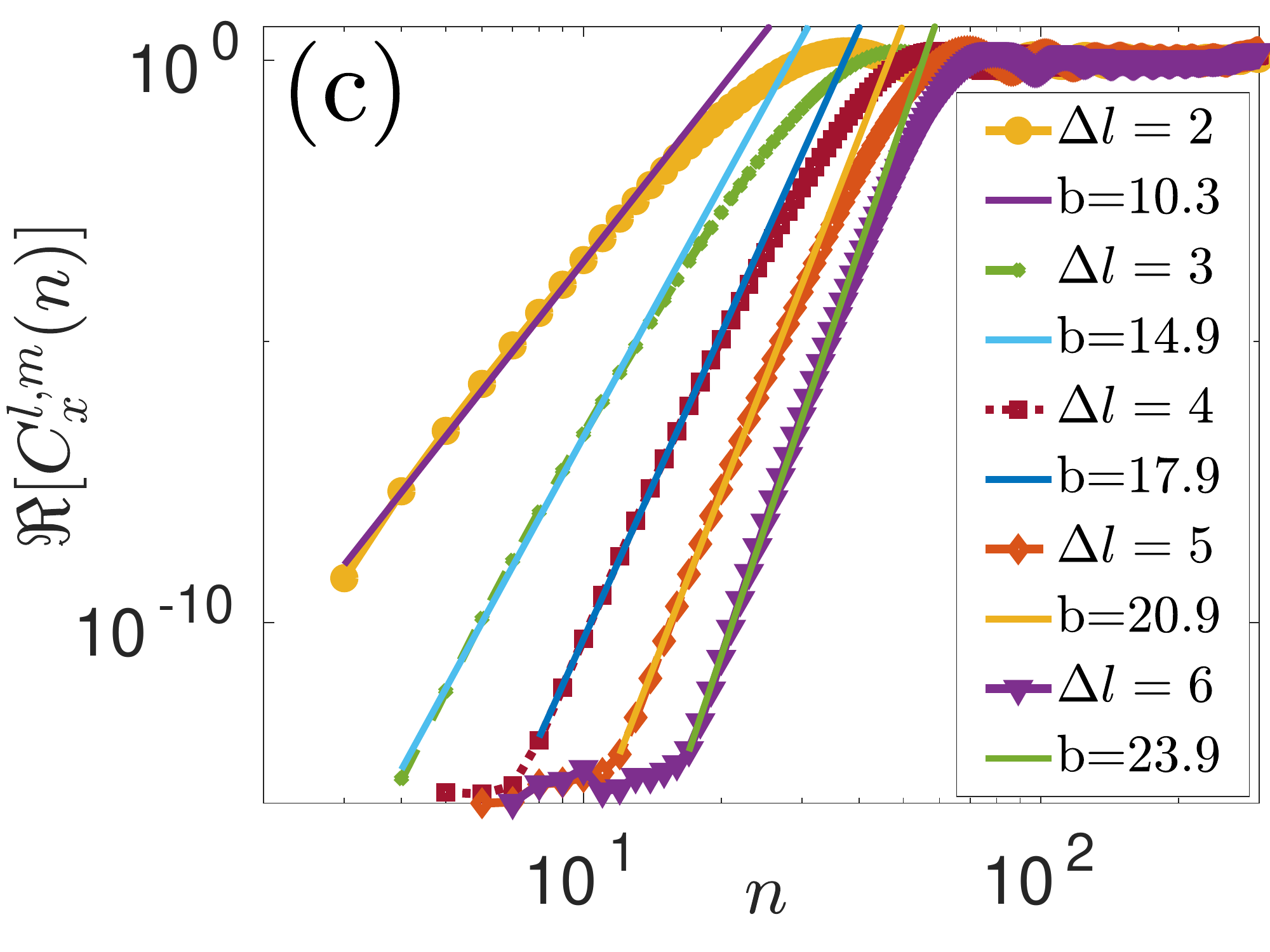}
  \end{subfigure}
  \begin{subfigure}{.30\textwidth}
  \includegraphics[width=.99\linewidth, height=.70\linewidth]{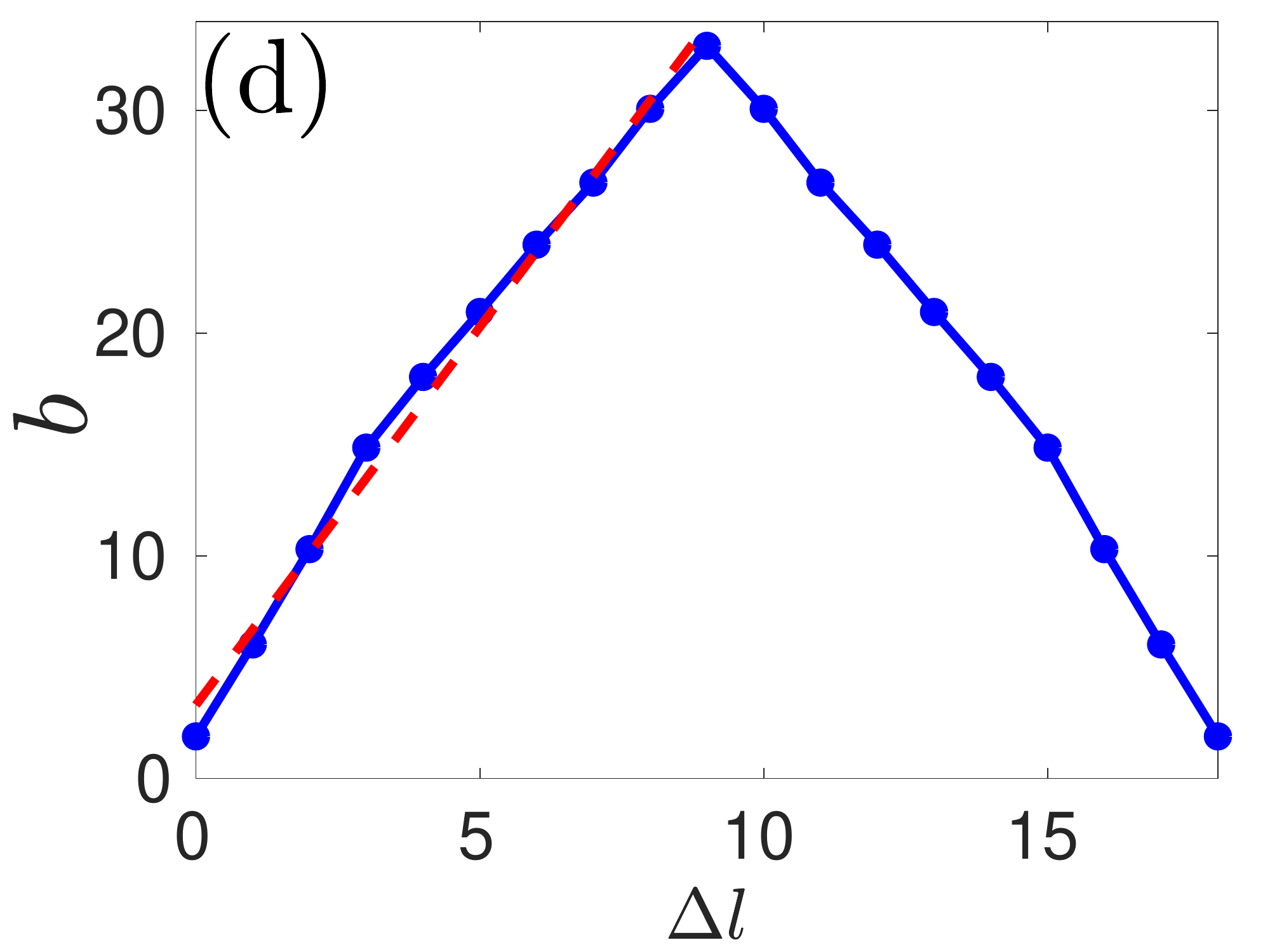}
  \end{subfigure}
  \begin{subfigure}{.30\textwidth} 
 \includegraphics[width=.99\linewidth, height=.70\linewidth]{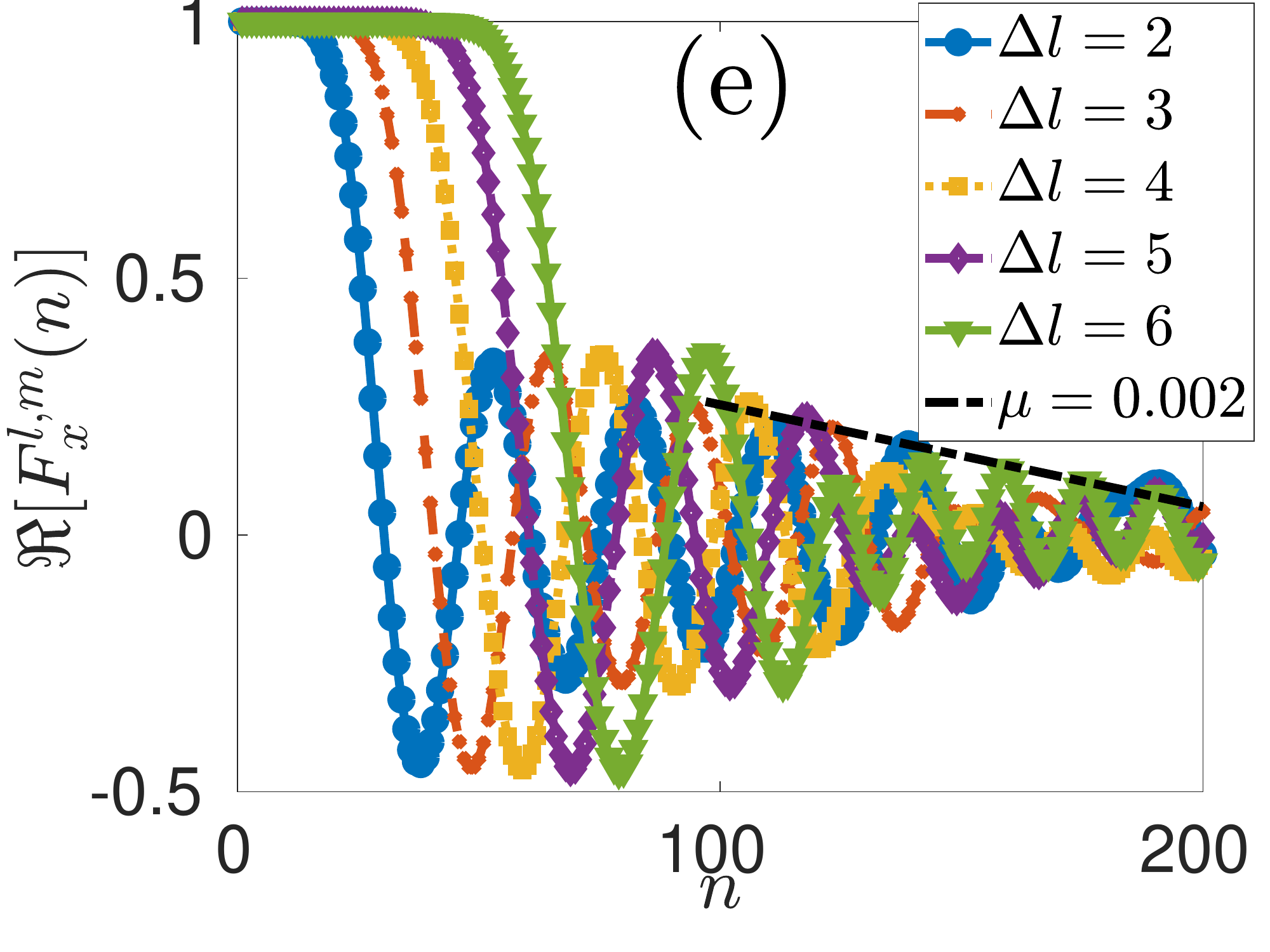}
 \end{subfigure}
 \caption{Integrable closed chain transverse Ising Floquet system with $J_x=1$ and $h_z=1$ of size $N=18$. (a) Behaviour of $LMOTOC$ with number of kicks $(n)$ with increasing value of Floquet periods from $\frac{7\epsilon}{2}$ to $\frac{11\epsilon}{2}$ (right to left) differing by $\frac{\epsilon}{2}$  and  $\Delta l=6$ ($\epsilon=\frac{\pi}{28})$. (b) $F^{l,m}_x(n)$ with number of kicks with increasing $\Delta l$ (left to right) and fixed Floquet period $\tau=6\epsilon/2$. (c)  $C^{l,m}_x(n)$ with number of kicks ($\log-\log$) with increasing $\Delta l$ (left to right) at fixed $\frac{\epsilon}{2}$. (d) Changing of power with $\Delta l$. (e)  $F^{l,m}_x(n)$ with number of kicks  
 at different increasing $\Delta l$ (left to right). Black dotted line represent the linear decreasing of maxima of saturation amplitude.} 
 \label{cf_LMOTOC_int}
\end{figure*}

\begin{figure*}[hbt!]
\begin{subfigure}{.30\textwidth} 
  \includegraphics[width=.99\linewidth, height=.70\linewidth]{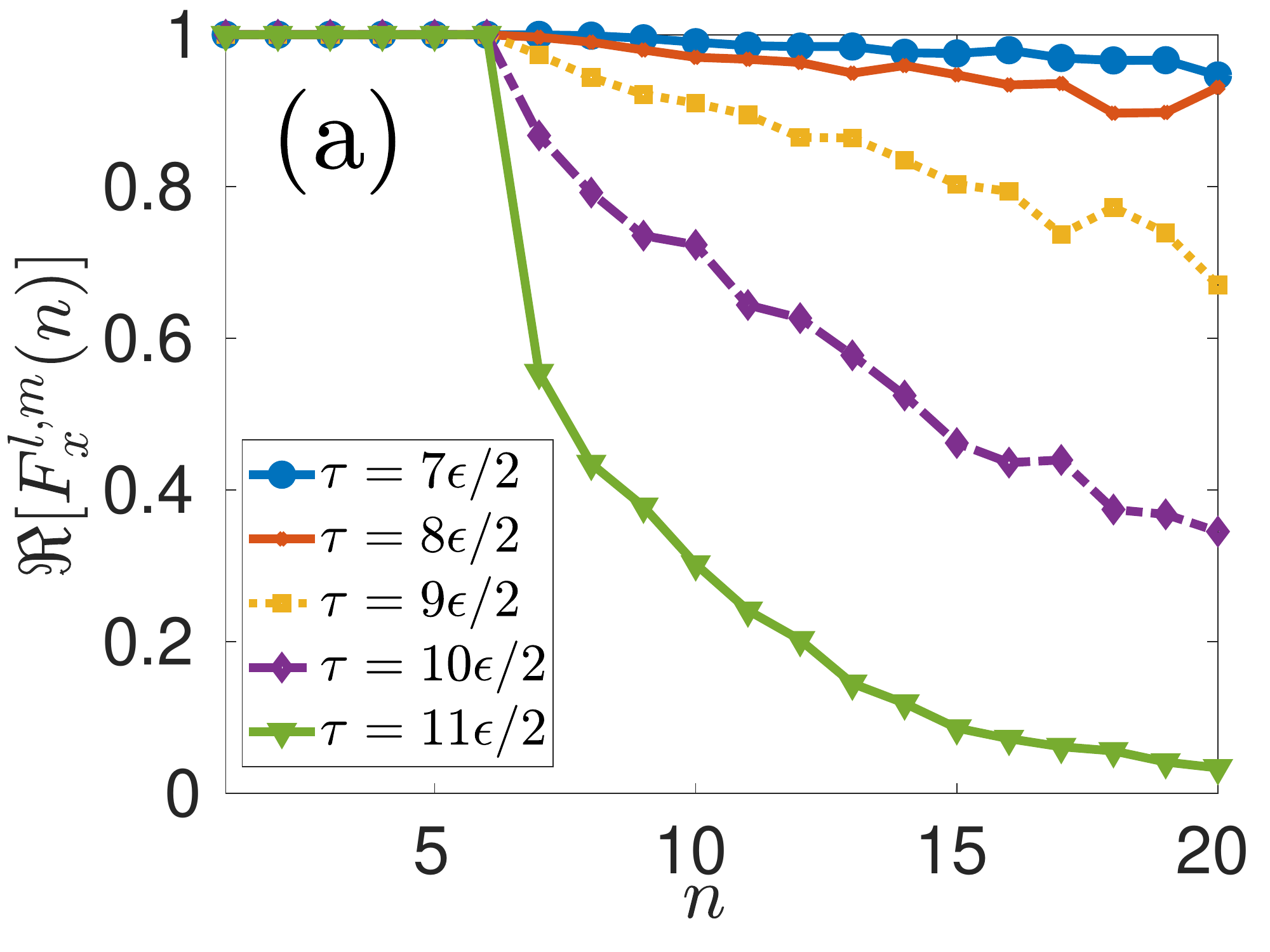}
 \end{subfigure}
 \begin{subfigure}{.30\textwidth} 
 \includegraphics[width=.99\linewidth, height=.70\linewidth]{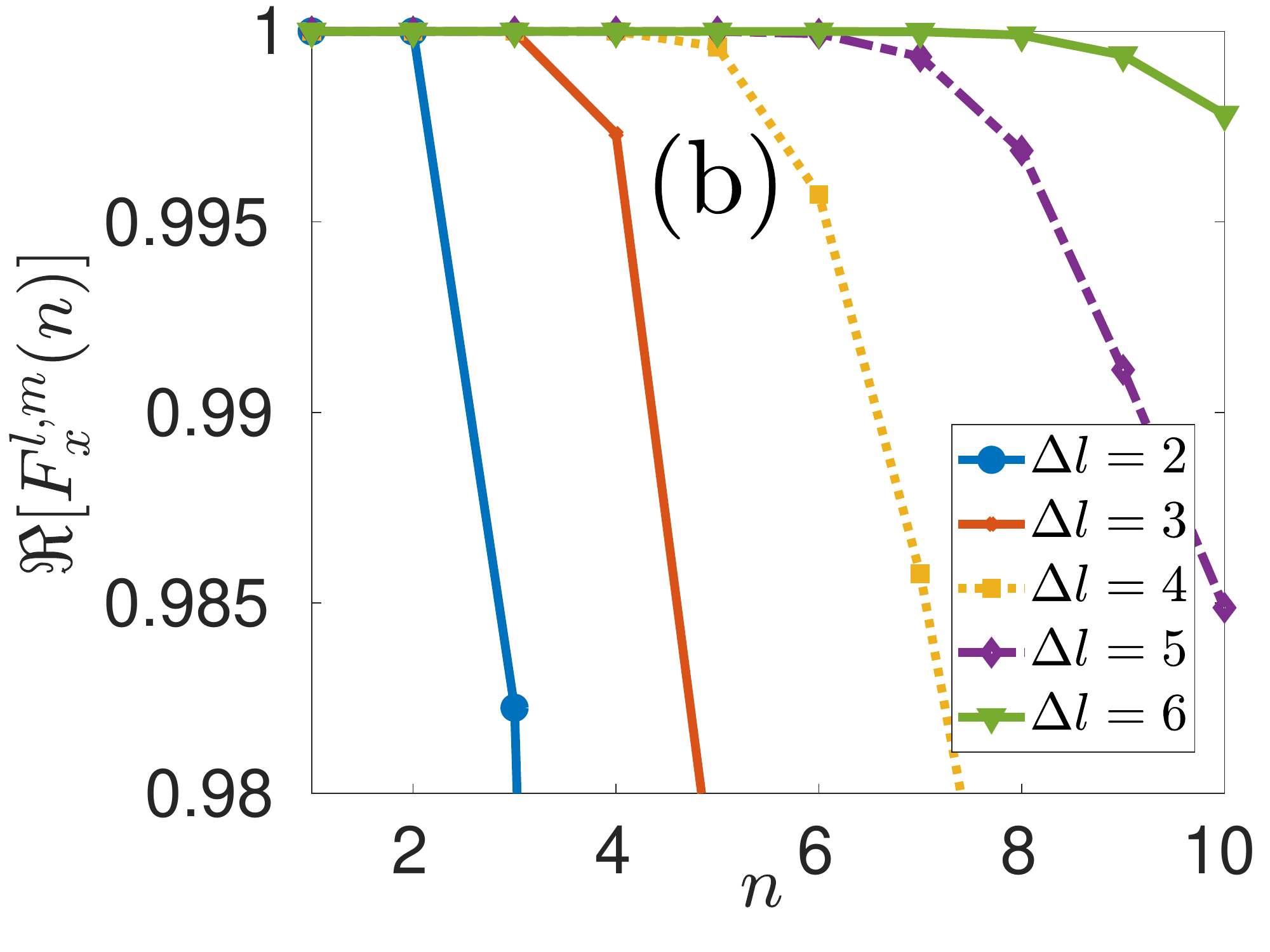}
 \end{subfigure}
 \begin{subfigure}{.30\textwidth} 
 \end{subfigure}
 \begin{subfigure}{.30\textwidth}
  \includegraphics[width=.99\linewidth, height=.70\linewidth]{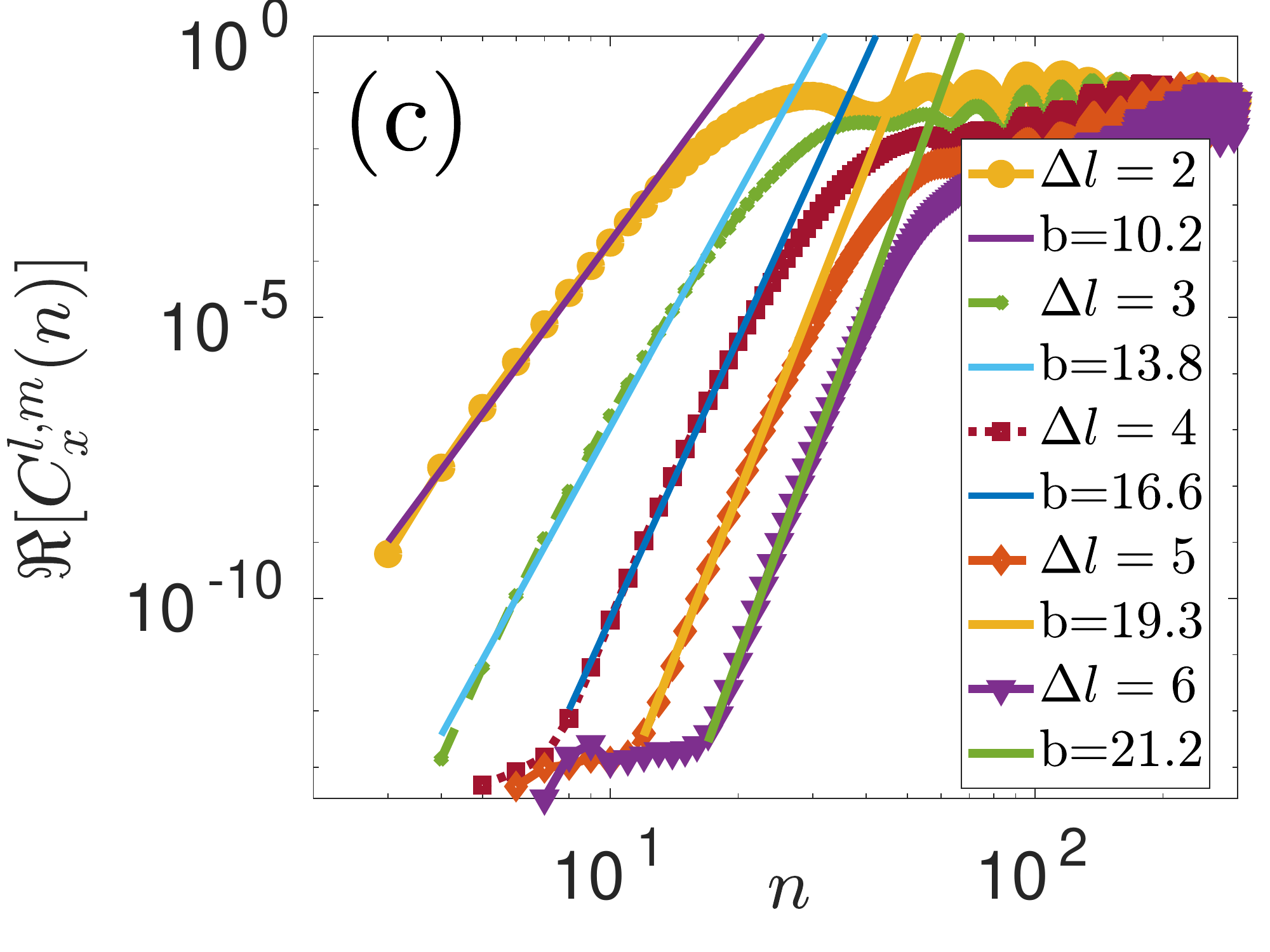}
  \end{subfigure}
  \begin{subfigure}{.30\textwidth}
  \includegraphics[width=.99\linewidth, height=.70\linewidth]{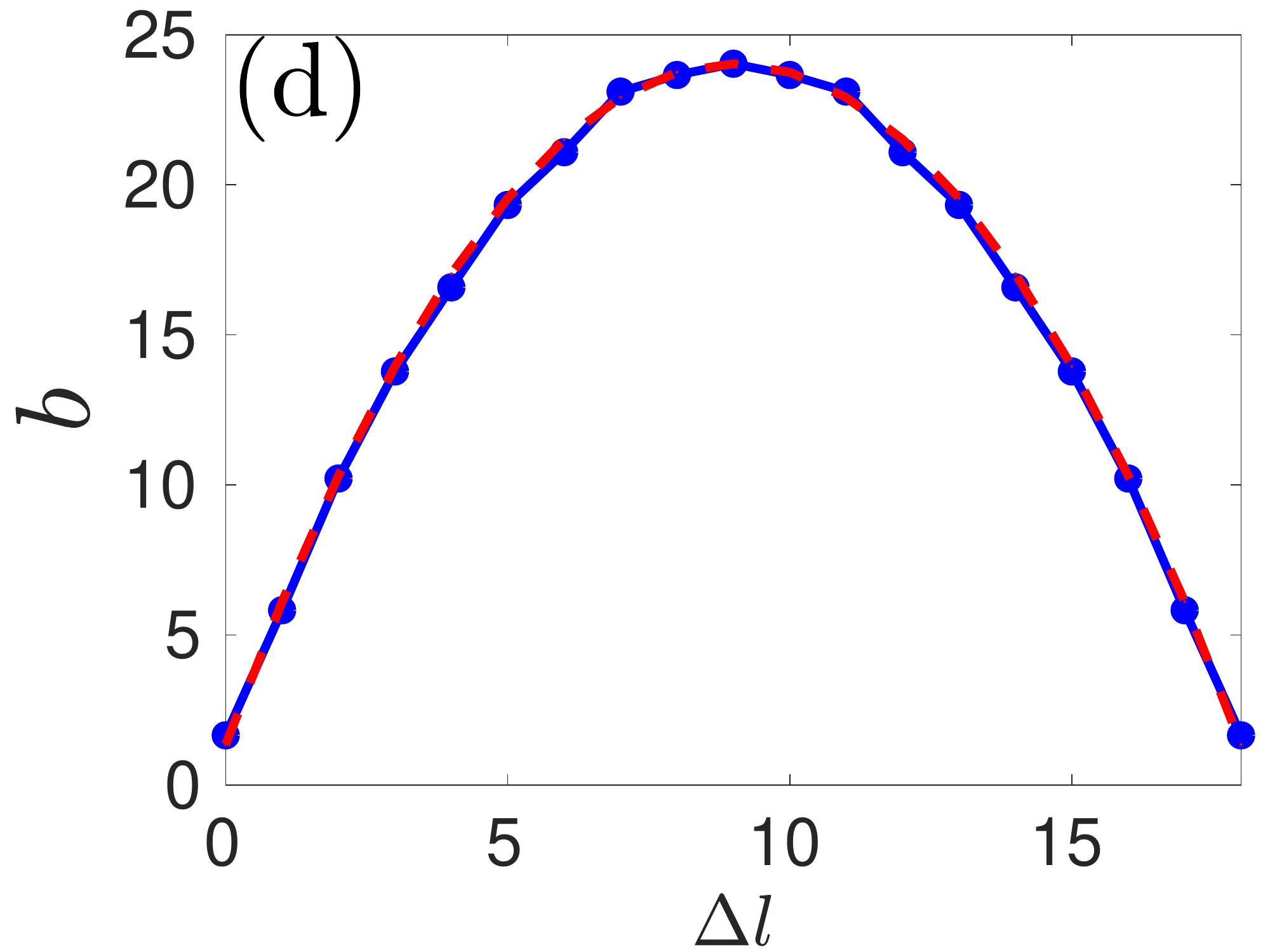}
  \end{subfigure}
  \begin{subfigure}{.30\textwidth}
  \includegraphics[width=.99\linewidth, height=.70\linewidth]{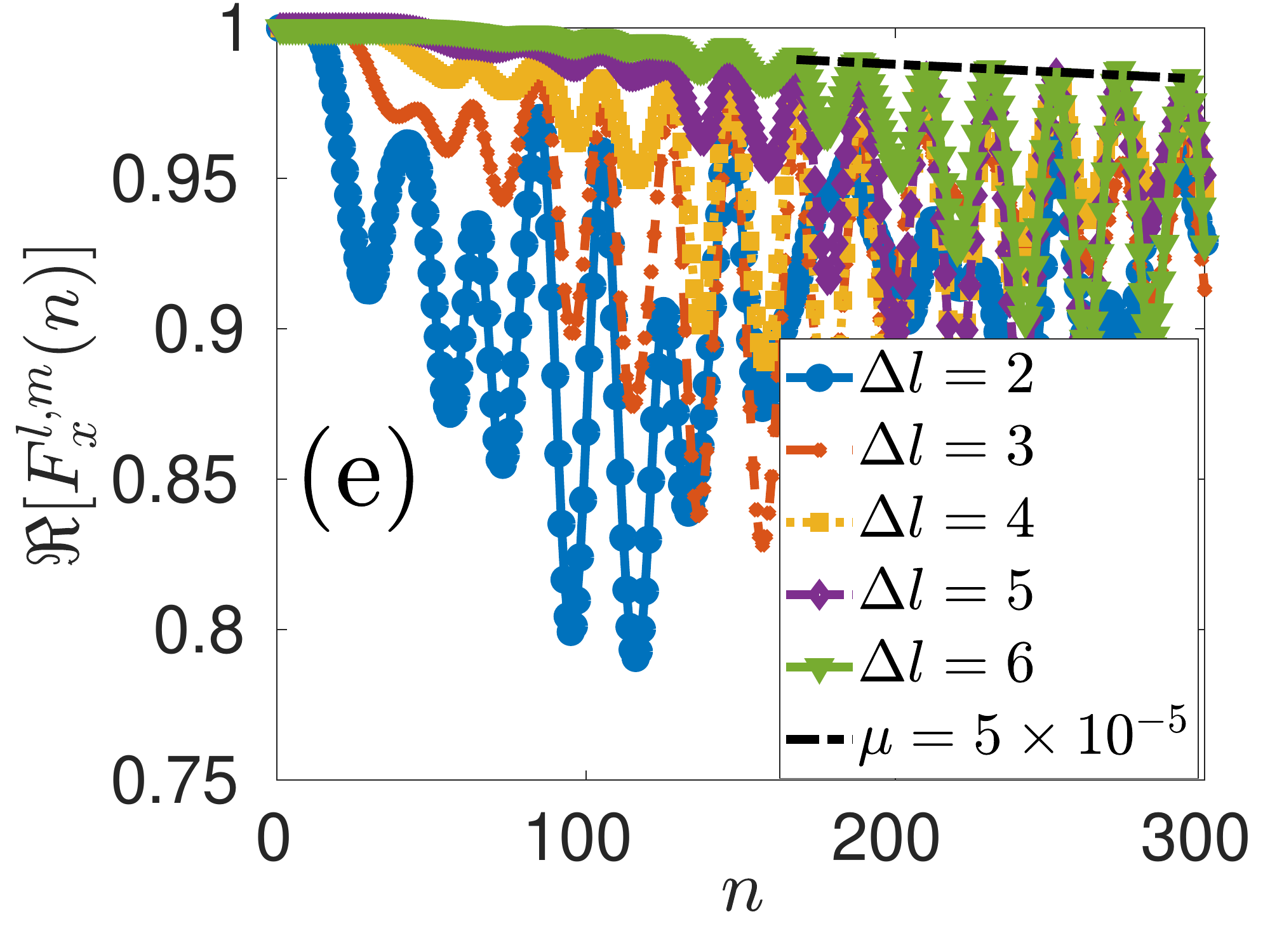}
  \end{subfigure}
 \caption{Non-integrable closed chain transverse Ising Floquet system  with $J_x=1$, $h_z=1$, and $h_x=1$ for $N=18$. (a) $LMOTOC$ with number of kicks $(n)$ by increasing value of Floquet period from $\frac{7\epsilon}{2}$  to $\frac{11\epsilon}{2}$ (right to left) differing by $\frac{\epsilon}{2}$  with fixed $\Delta l=6$ ($\epsilon=\frac{\pi}{28}$). (b)  $F^{l,m}_x(n)$ with increasing $\Delta l$ from $2$ to $6$ (left to right) and fixed period $\tau=\frac{6\epsilon}{2}$. (c)  $C^{l,m}_x(n)$ with number of kicks ($\log-\log$) by increasing $\Delta l$ from $2$  to $6$ (left to right) and fixed Floquet period $\tau=\frac{\epsilon}{2}$. (d) Changing of power with $\Delta l$. (e) $F^{l,m}_x(n)$ with number of kicks at different increasing $\Delta l$ (left to right). Black dotted line represent the linear decreasing of maxima of saturation amplitude.} 
 \label{cf_LMOTOC_nint}
\end{figure*}

\section{Results}
\label{result}
We analyse TMOTOC and LMOTOC for both $\mathcal{\hat{U}}_0$ and $\hat{\mathcal{U}}_x$ models in three regions as depicted in the Fig.~\ref{otocfig}. These are, namely:

\begin{enumerate}
    \item[i]) {\it Characteristic Region:} Both the observables $\hat W^l$ and $\hat V^m$ commute with each other till the characteristic time ($t_{\Delta l}$), which is defined as time after that $C^{l,m}_{z/x}(n)(F^{l,m}_{z/x}(n))$ is departed from zero (one). The Characteristic time depends upon the separation between the spins ($\Delta l=|l-m|$). As we increase the separation between the spins, the characteristic time increases and it is independent of the Floquet period and system size. 
    
    \item[ii]) {\it Dynamic Region:} After the characteristic time, $C^{l,m}_{z/x}(n)$ becomes nonzero.  In this dynamic region $C^{l,m}_{z/x}(n)$ increases rapidly.
    
    \item[iii]) {\it Near saturation Region:} After rapid growth, $C^{l,m}_{z/x}(n)$ starts to saturate to a finite value. However, the manner in which  $C^{l,m}_{z/x}(n)$ saturates follow some trend with an oscillating amplitude. Such trend we calculate by analysing behaviour of $\Re[F^{l,m}_{z/x}(n)]$ vs. $n$.
\end{enumerate}
\par
Let us begin with TMOTOC for $\mathcal{\hat{U}}_0$ system defined by Eq.(\ref{U0}). First, we  focus on the characteristic region of TMOTOC with increasing Floquet period. Let us consider an operator $\hat W$ located at site $l$ initially. One can see that the considered Floquet evolution increases the size of $\hat W$ at each Floquet step. In particular, the left end of the support of $\hat W(n)$ and the right end of the support of $\hat W(n)$ will increase by one for each Floquet step. We, therefore, can see that  $F_z^{l,m}(n)=1$ if $n < |l - m|$. However, once $n \geq |l - m|$, $F_z^{l,m}(n)$ will start to deviate from $1$. 
\par
 Fig.~\ref{cf_TMOTOC_int}(a) is the behaviour of $F_z^{l,m}(n)$ with increasing Floquet period and fixed $\Delta l=6$. One can see from Fig.~\ref{cf_TMOTOC_int}(a) that $F_z^{l,m}(n)$ start to deviate at $\Delta l^{\rm th}(=6^{\rm th})$ kick for all Floquet period $(\tau)$. This characteristic time is independent
of the Floquet period the system size $N$ (We have checked till $N=50$). For a fixed Floquet period $\tau$ we can see the behaviour of $\Re[F^{l,m}_z(n)]$ with the number of kicks and see the dependence of $t_{\Delta l}$ on $\Delta l$. In Fig.~\ref{cf_TMOTOC_int}(b), for 
$\tau=\frac{6\epsilon}{2}$, we show $\Re[F^{l,m}_z(n)]$  vs. the number of kicks by changing the separation between the observables $\Delta l=|l-m|$. We see that increasing the separation between the spins, increases the characteristic time for the TMOTOC case  and number of kicks required to deviate from unity is equal to the seperation between the observables ($n=\Delta l$).  The growth of TMOTOC in the dynamic region follows a power-law. The exponent of the power-law increases with increasing the separation between the local spin observables in a systematic manner [Fig.~\ref{cf_TMOTOC_int}(c)]. The exponent increases, reaches maximum at $\Delta l=\frac{N}{2}$, and further decreases with increasing the distance between the spins [Fig.~\ref{cf_TMOTOC_int}(d)]. The exponent of the power-law can be expressed as a triangular function:
\begin{eqnarray}
b \approx b_{\rm max}-\kappa\Big\vert\frac{N}{2}-\Delta l\Big\vert,  \quad\quad\quad 1\leq\Delta l\leq N-1.
\label{bformula}
\end{eqnarray}
where, the constants $\kappa=3.2$, $b_{\rm max}=29$ and $b_0=1.7$.   Eq.~(\ref{bformula}) shows dependence of the exponent of power-law with increasing the separation between the observables. It is symmetric about $\Delta l=\frac{N}{2}$ because of the periodic boundary condition of the spin chains. 
$\Re[F^{l,m}_z(n)]$ revives back to unity after a few kicks,in the saturation region. Revival  time has nontrivial dependence on $n$ and $\Delta l$    [Fig.~\ref{cf_TMOTOC_int}(e)]. 
The TMOTOC extracted from the analytical expression Eq. (\ref{OTOCz_gene}) in characteristic, dynamic and saturation regions can be summarised as  
\begin{equation}
\label{C_int_tmotoc}
C_z^{l,m}(n) \approx \left\{
\begin{aligned}
 0, & \quad\quad\quad n\tau< t_{\Delta l},\\
 (n\tau)^{\kappa\Delta l+2}, & \quad\quad\quad  t_{\Delta l}<n\tau<t_s,\\
 {\rm revived \quad back}, & \quad\quad\quad   t_s<n\tau.
\end{aligned} \right.
\end{equation}
 In the above expression $t_{\Delta l}$ is characteristic time  and $t_s$ is the time at which TMOTOC starts saturating.  Dynamic region of TMOTOC decreases with increasing the Floquet period $\tau$ as shown in Fig~\ref{cf_TMOTOC}(a-e). In general the dependence on $\tau$ is such that we can define $C_z^{l,m}\propto (n\tau)^{\kappa\Delta l+2}$ in dynamic region. 
\par
Now, we use the nonitegrable $\hat{\mathcal{U}}_x$ model given by Eq. (\ref{Ux}) and analyze the TMOTOC. Fig.~\ref{cf_TMOTOC_nint}(a) shows the behavior of $F^{l,m}_z(n)$ for varying $\tau$ and fixed $\Delta l=6$. From Fig.~\ref{cf_TMOTOC_nint}(a), one can see that number of kicks required for $F_z^{l,m}(n)$ depart from unity is equal to the separation between the observables ($n=\vert l-m\vert$). Hence characteristic time does not depend on the Floquet periods. Let us explore the behaviour of TMOTOC with distance between the spins for a fixed $\tau$ (say $\tau=\frac{6\epsilon}{2}$) and increase the separation between the spins $\Delta l$. As $\Delta l$ increases, the characteristic time $(t_{\Delta l})$ increases  in such a way that $n=\Delta l$ [Fig.~\ref{cf_TMOTOC_nint}(b)]. Dynamic region of TMOTOC for the nonintegrable is again showing power-law and the exponent of the power-law $(b)$ depends on $\Delta l$ [Fig.~\ref{cf_TMOTOC_nint}(c)]. $b$ increases with increasing $\Delta l$ and reaches a maximum $(b_{\rm max})$  at $\Delta l=\frac{N}{2}$ and afterwards decreases symmetrically with increasing $\Delta l$ before coming down to $b_1$ at $\Delta l=N-1$. Since, we consider the periodic boundary condition, the exponent of the power-law is symmetric about $\Delta l=\frac{N}{2}$ [Fig.~\ref{cf_TMOTOC_nint}(d)]. In a mathematical form we can express $b$, approximately, by Eq.(\ref{bformula}) with $\kappa =3.2$, $b_{\rm max}=28$ and $b_{min}=1.78$. Saturation of $\Re[F_z^{l,m}(n)]$ in this nonintegrable model is following a linear decaying behaviour with a very small slope  for all $\Delta l$ [Fig.~\ref{cf_TMOTOC_nint}(e)].
 TMOTOC for $\hat{\mathcal{U}}_x$ model in all the regions is summed up as
 \begin{equation}
 \label{C_tmotoc_nint}
C_z^{l,m}(n) \approx \left\{
\begin{aligned}
 0, & \quad\quad\quad n\tau< t_{\Delta l},\\
 (n\tau)^{\kappa\Delta l+1}, & \quad\quad\quad  t_{\Delta l}<n\tau<t_s,\\
 1- \mu n, & \quad\quad\quad   t_s<n\tau.
\end{aligned} \right.
\end{equation}
where $\mu=0.002$ and $\kappa=3.2$. We calculate the exponent of the power-law by using the HBC formula for $\Delta l=1,2$ and find approximate matches with the exponent of the power-law in the dynamic region of Eq.~(\ref{C_tmotoc_nint}).  Detailed calculation is given in the Appendix \ref{HBC_TMOTOC}. 
 \par
 Now we focus on the LMOTOC for the  $\hat{\mathcal{U}}_0$ model which shows a similarity with TMOTOC for the same model. 
 Fig.~\ref{cf_LMOTOC_int}(a)  is the behaviour of LMOTOC at different Floqeut periods and fixed $\Delta l=6$. In the LMOTOC, number of kicks required to deviate from unity is $n=\Delta l+1$. In comparison with TMOTOC, LMOTOC required one more kick to deviate $F_x^{l,m}(n)$  from unity because $\hat \sigma_x^l$ (using Baker–Campbell–Hausdorff formula) provide spreading terms after the first kick. 
  Hence, characteristic time does not depend on the Floquet period,  however, depends on the speration between the observables in such a way that characteristic time increases linearly with increasing the separation between the observables ($n=\Delta l+1)$ [Fig.~\ref{cf_LMOTOC_int}(b)] .  
In the dynamic region of LMOTOC at small Floquet period, similar to the TMOTOC case, we get a power-law behaviour. The exponent of the power-law increase with $\Delta l$ in the same manner as in TMOTOC case [Fig.~\ref{cf_LMOTOC_int}(c)]. We can approximate the exponent with $\Delta l$ by Eq.(\ref{bformula}) with $\kappa=3.4$, $b_{\rm max}=32.9$ and $b_{0}=1.9$ [Fig.~\ref{cf_LMOTOC_int}(d)].   
 Saturation region of LMOTOC for  $\hat{\mathcal{U}}_0$ shows oscillating behavior. The envelope of the oscillation decays linearly with a constant slope for all $\Delta l$ [Fig.~\ref{cf_LMOTOC_int}(e)]. This behaviour is a contrast to the saturation region of TMOTOC for $\hat{\mathcal{U}}_0$ which displays a revival to early-time behaviour.  All the regions of LMOTOC  for  $\hat{\mathcal{U}}_0$ can be encapsulated as
\begin{equation}
\label{C_lmotoc_int}
C_x^{l,m}(n) \approx \left\{
\begin{aligned}
 0, & \quad\quad\quad n\tau< t_{\Delta l},\\
 (n\tau)^{\kappa\Delta l+6}, & \quad\quad\quad   t_{\Delta l}<n\tau<t_s,\\
 1- \mu n, & \quad\quad\quad   t_s<n\tau.
\end{aligned} \right.
\end{equation}
\par
Finally, we consider $\hat{\mathcal{U}}_x$ model for LMOTOC calculations. We get a similar behaviour in characteristic regime as that for LMOTOC with $\hat{\mathcal{U}}_0$ model [Fig.~\ref{cf_LMOTOC_nint}(a) and (b)]. 
In the dynamic region, the growth is again a power-law, and the exponent increase with $\Delta l$ [Fig.~\ref{cf_LMOTOC_nint}(c)] but the trend is a bit different than the previous cases. Unlike the previous cases, we see a quadratic increase of the exponent by increasing $\Delta l$, till a maximum is reached. After the maximum $b_{\rm max}$ at $\Delta l=\frac{N}{2}$, we see a symmetric decrease in the exponent till $\Delta l=N$ [Fig.~\ref{cf_LMOTOC_nint}(d)]. We approximate $b$ as follows:
\begin{equation}
\label{bformula_lmotoc}
b \approx \Big(b_{\rm max}-\lambda\Big\vert\frac{N}{2}-\Delta l\Big\vert^2\Big).\quad\quad\quad 0\leq \Delta l \leq N
\end{equation}
Where $\lambda=2.8$, $b_{\rm max}=24.0$ and $b_{0}=1.7$.  Eq.~(\ref{bformula_lmotoc}) describes the variation of exponent of power-law with increasing the separation between the observables. It is parabolic form with vertex at $\frac{N}{2}$ and also symmetric about $\Delta l=\frac{N}{2}$ because of closed chain consideration. We calculate the exponent of the power-law by using the HBC formula for $\Delta l=1$ and find that exponent approximately matches the Eq.~(\ref{bformula_lmotoc}). Detailed calculation is mentioned in Appendix \ref{HBC_LMOTOC}.  
Saturation of LMOTOC for nonintegrable case is  oscillating and the maxima of the oscillation decreases linearly (with a very small slope $ \mu=10^{-5}$, and same for all $\Delta l$ ) with the number of kicks [Fig.~\ref{cf_LMOTOC_nint}(e)]. 
The complete region of LMOTOC for $\hat{\mathcal{U}}_x$ system is given as 
\begin{equation}
\label{C_lmotoc_nint}
C_x^{l,m}(n) \approx \left\{
\begin{aligned}
 0, & \quad\quad\quad n\tau< t_{\Delta l},\\
 (n\tau)^{\lambda(\Delta l)^2}, & \quad\quad\quad   t_{\Delta l}<n\tau<t_s,\\
 1- \mu n, & \quad\quad\quad   t_s<n\tau.
\end{aligned} \right.
\end{equation}
In a nutshell, we see that the characteristic regions of LMOTOCs have similar behaviour for $\hat{\mathcal{U}}_0$ and $\hat{\mathcal{U}}_x$ systems. In both  cases, the commutator propagation varies with $\tau$ in a similar way. But the dynamic region displays a contrast between $\hat{\mathcal{U}}_0$ and $\hat{\mathcal{U}}_x$. In the integrable case, the exponent of the power-law increases linearly with $\Delta l$, but in the nonintegrable exponent, we see a quadratic growth of power-law with $\Delta l$. In the saturation region, both are oscillating and the envelope decreases with different rates.
\par
In this manuscript we considered single spins as observables in our calculation of OTOCs. The experimental procedure of the calculation of OTOC  using single spin observables and  initial product state has been done in Ref.~\cite{joshi2020quantum}.  Implementation of the unitary operator on observable $\hat W^l$  \big[$\hat W^l(n)=(\mathcal{\hat U}_x^{\dagger})^{n}\hat W^l(0)(\mathcal{\hat U}_x)^n$\big] followed by perturbation of observable $\hat V^m$ is discussed in the Ref.~\cite{garttner2017measuring}. The OTOC is obtained by measuring the expectation value of the observable $(\mathcal{\hat U}_x^{\dagger})^n \hat W^l(0)(\mathcal{\hat U}_x)^n \hat V^m (\mathcal{\hat U}_x^{\dagger })^n \hat W^l(0)(\mathcal{\hat U}_x)^n \hat V^m$ \cite{joshi2020quantum}. Therefore, LMOTOCs and TOMOTOCs can be calculated experimentally.

\section{Conclusion}
\label{conclusion}
We studied the behaviour of TMOTOC and LMOTOC comprehensively using  $\mathcal{\hat{U}}_0$ and $\mathcal{\hat{U}}_x$ systems. We divided LMOTOC and TMOTOC into three distinct regimes: characteristic-time regime, dynamic-time regime, and saturation-time regime. 
\par
Characteristic times of TMOTOC and LMOTOC are independent of the integrability of the system. It also independent of the Floquet period and system size, however, it depends on the separation between the observables. Number of kicks required for the deviation of $F^{l,m}$ from unity is equal to the numerical value of the separation between the observables in the case of TMOTOC, howerver, one extra kick is required in the case of LMOTOC. Behaviour of the dynamic region is also independent of the integrability of the system. In both systems, $\mathcal{\hat U}_0$ and $ \mathcal{\hat U}_x$, LMOTOC and TMOTOC show the power-law growth. There is no signature of Lyapunov exponent. This power-law growth depends on the separation between the spins and the Floquet period. The rate of change of exponent with respect to the separation between the spins is independent of the integrability of the system in the TMOTOC, however, we see a dependence in the case of LMOTOC. In TMOTOC for both the systems $\mathcal{\hat U}_0$, and $\mathcal{\hat U}_x$, the exponent varies as a  triangular function.  In the case of LMOTOC, we see a triangular function with linear increase/decrease for $\mathcal{\hat U}_0$ system but a quadratic increase/decrease for $\mathcal{\hat U}_x$ system. Saturation region of TMOTOC is different in both systems:  $\mathcal{\hat U}_0$ system revives back but $\mathcal{\hat U}_x$ system decays linearly.   Saturation behavior of LMOTOC shows the oscillating decay with envelop decaying linearly in both  systems. Saturation of TMOTOC and LMOTOC are independent of $\Delta l$. 
\bibliographystyle{apsrev4-1}
\bibliography{scramb}
\appendix

\begin{widetext}
Calculation of TMOTOC in the non-integrable $\mathcal{\hat U}_x$ system using random state
\label{appendix1}

 If $\hat V$ and $\hat W$ are two Hermitian operators that are localized on different positions $l$ and $m$, respectively, the OTOC  \cite{larkin1969quasiclassical} is given as:
\begin{eqnarray}
C^{l,m}(n)=-\frac{1}{2}\mbox{Tr} \left([\hat W^l(n), \hat V^m]^2\right),
\label{Cn1}
\end{eqnarray}
which is a measure of the noncommutativity of two operators $\hat W^l$ and $\hat V^m$. These are infinite temperature quantities and involves the entire spectrum of $2^{N}$ states. One can use the trick for evaluating OTOC by employing Haar random states of $2^N$ dimensions ($\vert\Psi_R \rangle$) and calculate expectation value over $\vert\Psi_R \rangle$. OTOC will be 
\begin{eqnarray}
C^{l,m}(n)=-2^{N-1} \langle \Psi_R|[\hat W^l(n), \hat V^m]^2\vert\Psi_R\rangle,
\end{eqnarray}
Since, the behaviour of OTOC is similar in both the cases either taking random states or special initial states ($\vert\phi\rangle$ and $\vert\psi\rangle$ accordingly). So we consider special initial states and OTOC will be 
\begin{eqnarray}
C^{l,m}(n)=-2^{N-1} \langle \psi/\phi|[\hat W^l(n), \hat V^m]^2\vert\psi/\phi\rangle,
\end{eqnarray} 

\begin{figure*}[hbt!]

\begin{subfigure}{.30\textwidth}
  \includegraphics[width=.99\linewidth, height=.70\linewidth]{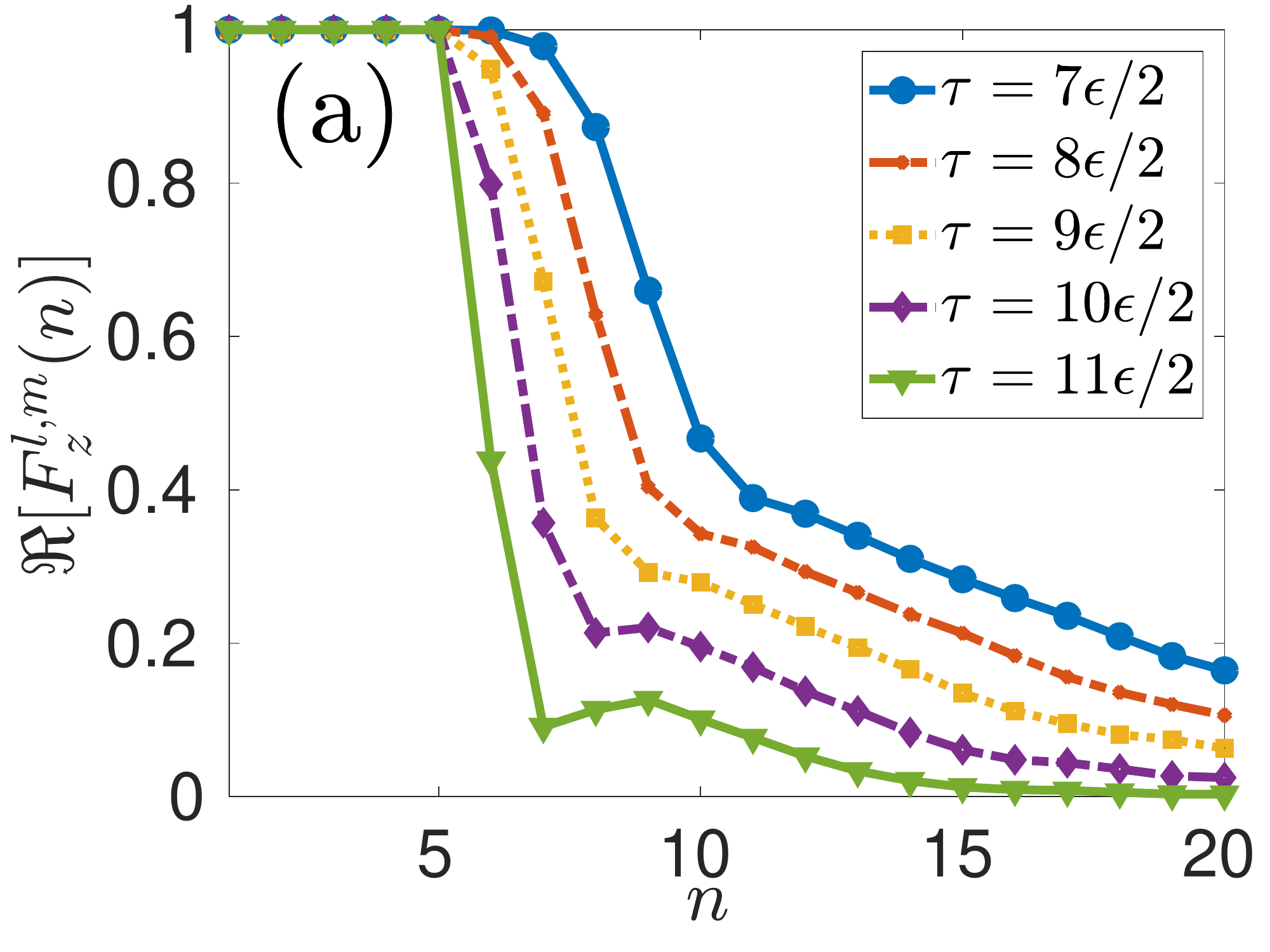}
  \end{subfigure}
  \begin{subfigure}{.30\textwidth}
  \includegraphics[width=.99\linewidth, height=.70\linewidth]{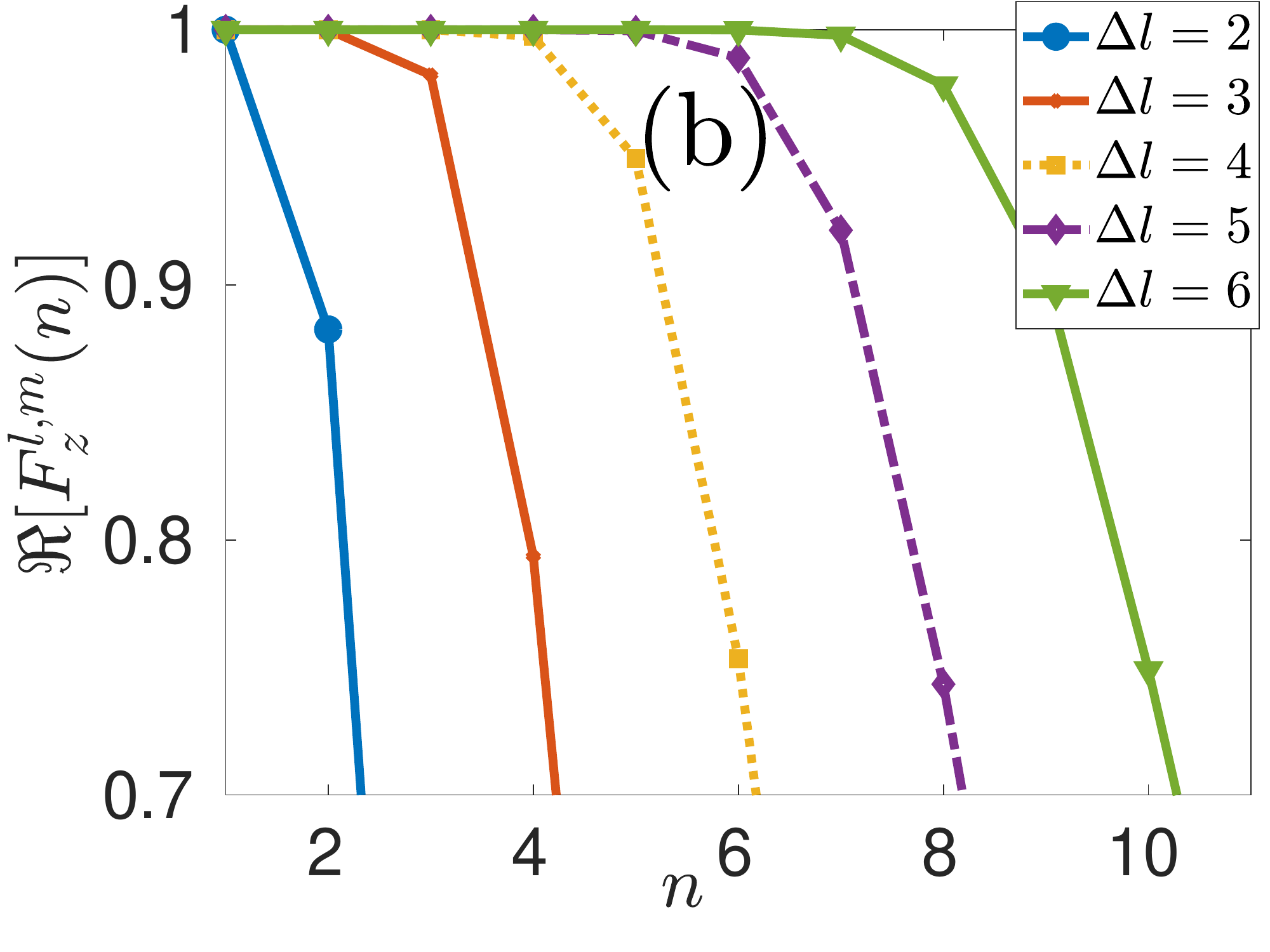}
  \end{subfigure}
\begin{subfigure}{.30\textwidth}
  \includegraphics[width=.99\linewidth, height=.70\linewidth]{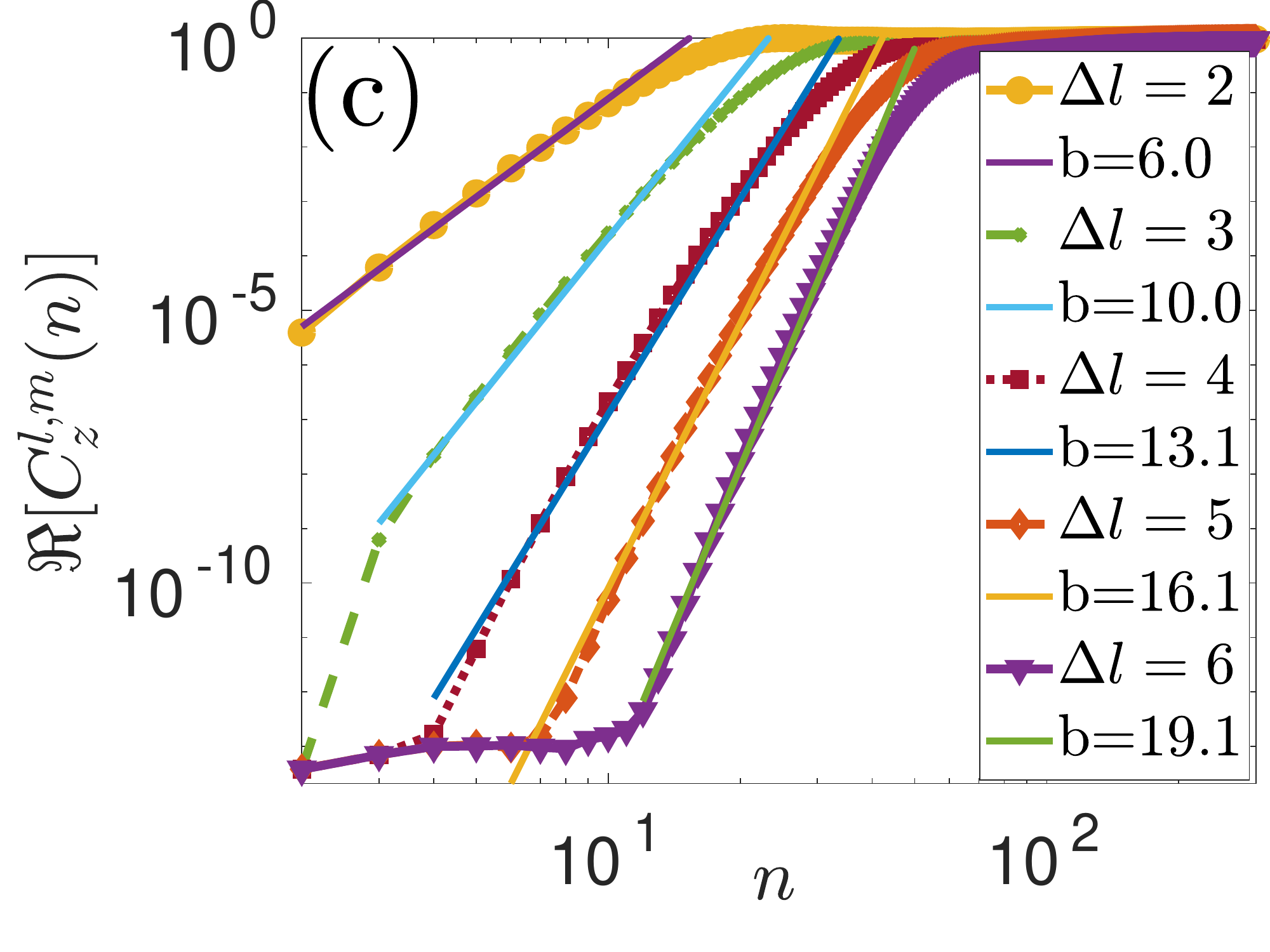}
  \end{subfigure}
  \begin{subfigure}{.30\textwidth}
  \includegraphics[width=.99\linewidth, height=.70\linewidth]{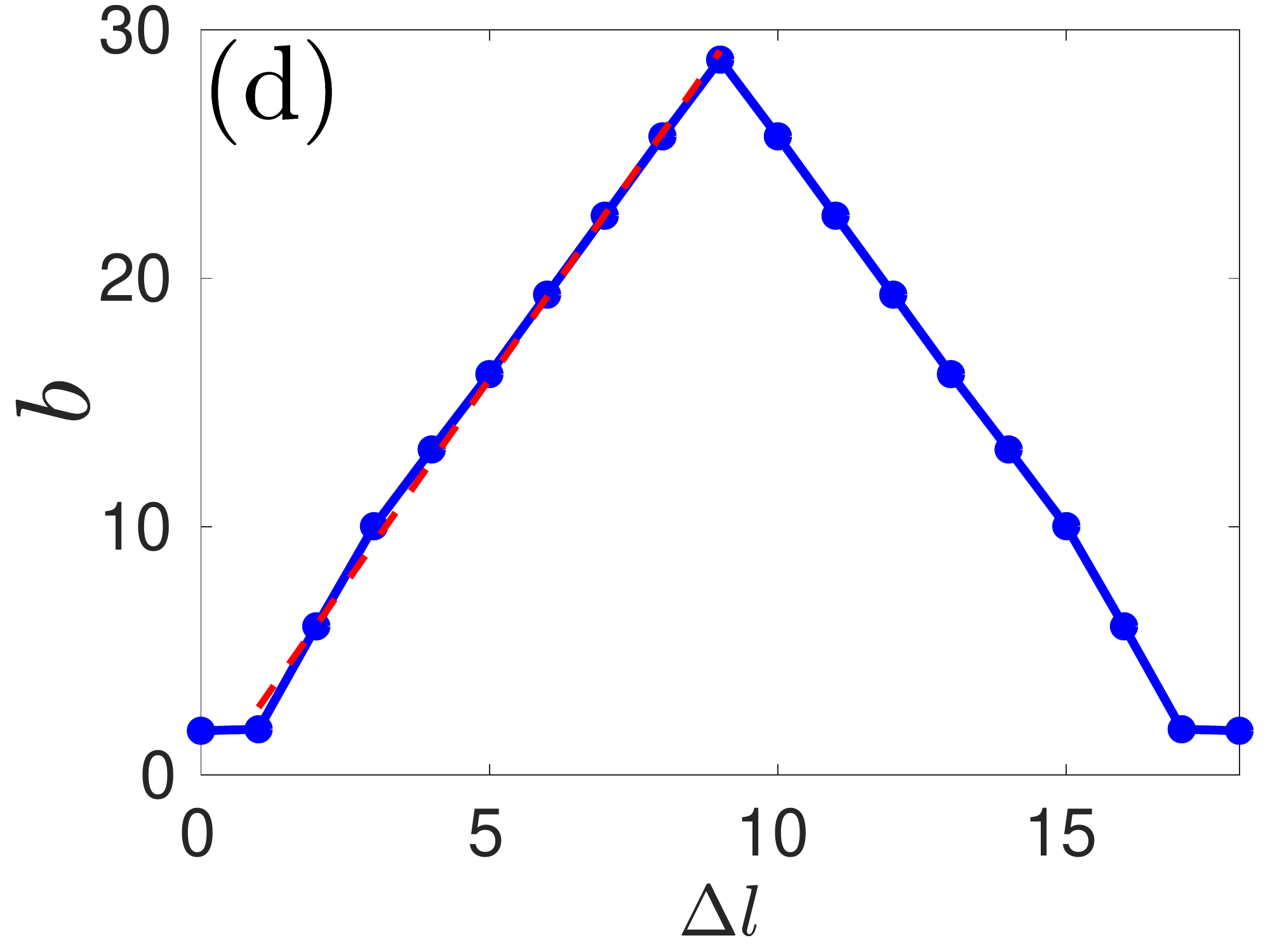}
  \end{subfigure}
  \begin{subfigure}{.30\textwidth}
  \includegraphics[width=.99\linewidth, height=.70\linewidth]{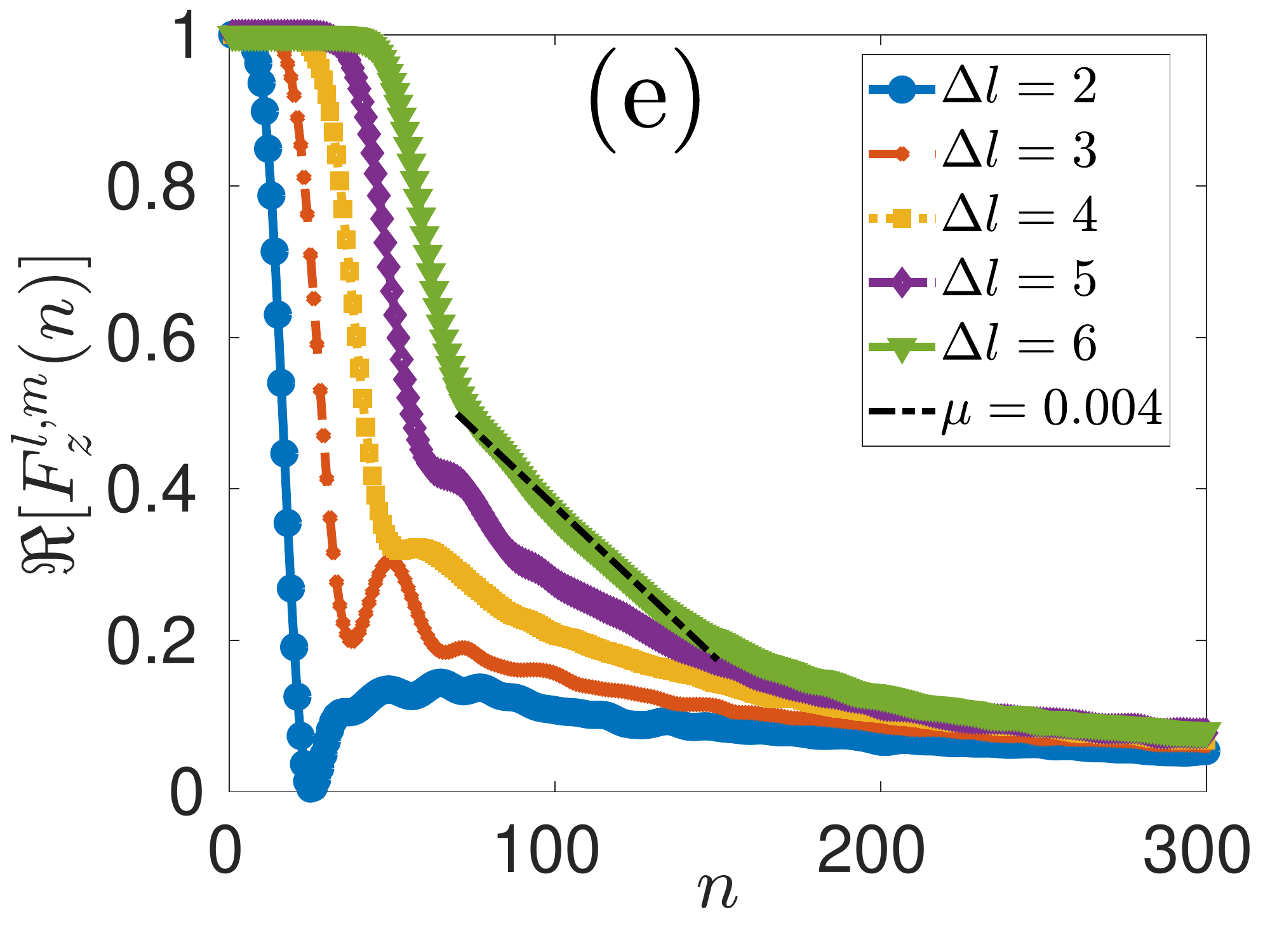}
  \end{subfigure}
 \caption{Nonintegrable closed chain transverse Ising Floquet system  with $J_x=1$, $h_z=1$, and $h_x=1$ of size $N=18$. (a) Behaviour of $TMOTOC$ with number of kicks $(n)$ by increasing Floquet period from  $\frac{7\epsilon}{2}$ to $\frac{11\epsilon}{2}$ (right to left) differing by  $\frac{\epsilon}{2}$ with fixed $\Delta l=6$ ($\epsilon=\frac{\pi}{28})$.  (b)  Initial region of $F^{l,m}_z$ with number of kicks and increasing $\Delta l$ (left to right) at fixed Floquet period  $\tau=6\epsilon/2$. (c)   $C^{l,m}_z$ with number of kicks ($\log-\log$) with increasing $\Delta l$ (left to right) at fixed $\tau=\frac{\epsilon}{2}$.  (d) Changing of exponent of power-law with $\Delta l$. (e) Saturation of $F^{l,m}_z$ with number of kicks with incresing $\Delta l$ (left to right). }
\label{rand_cf_TMOTOC_nint}
\end{figure*}
Fig.~(\ref{rand_cf_TMOTOC_nint}) is the behaviour of TMOTOC in the nonintegrable $\mathcal{\hat U}_x$ system using random initial state $(\psi_R)$ drawn form the Harr measure. Characteristic time is independent of the Floqeut period [Fig.~\ref{rand_cf_TMOTOC_nint}(a)] and it depends on the separation between the observables. Number of kicks required to depart from unity is equal to separation between the observables [Fig.~\ref{rand_cf_TMOTOC_nint}(b)]. Dynamic region of TMOTOC for the nonintegrable is showing a power-law [Fig.~\ref{rand_cf_TMOTOC_nint}(c)] that is approximately similar to the [Fig.~\ref{cf_TMOTOC_nint}(c)]. The exponent of the power-law $(b)$ depends on $\Delta l$ [Fig.~\ref{rand_cf_TMOTOC_nint}(d)] and its behaviour is approximately similar as Fig.~\ref{cf_TMOTOC_nint}(d). 
Saturation of $\Re[F_z^{l,m}(n)]$ is following a linear decaying behaviour with a very small slope ($0.004$) for all $\Delta l$ [Fig.~\ref{rand_cf_TMOTOC_nint}(e)]. There is very small oscillation in comparison of Fig.~\ref{cf_TMOTOC_nint}(e).

\section{Time evolution of TMOTOC}
 \label{HBC_TMOTOC}
  The Heisenberg evolution of an operator $\hat W(t)$ can be expanded using the Hausdorff-Baker-Campbel (HBC) formula as
\begin{equation}
\label{HBC}
\hat W(t)=\sum_{p=0}^{\infty}
\frac{(i t)^p}{p!}
[\hat H,[ \hat H,\cdots^{{\rm p~times}},[\hat H,\hat W]]]. 
\end{equation}
If $\hat W=\hat \sigma^{z/x}_l$, the HBC formula captures the spread of the operator over the spin sites and how it becomes more complex as time increases. Furthermore, direct replacement of Eq.~\ref{HBC} in Eq.~\ref{gene_OTOC}
highlights the fact that the short-time growth is characterized
by the smallest $p$ on which
\begin{equation}
    [\hat H,[ \hat H,\cdots^{{\rm p~times}},[\hat H,\hat \sigma^{x/z}_l]], \hat \sigma^{x/z}_m] \neq 0,
\end{equation}
due to the time factor $t^n$ that weights the terms in the expansion. We remark that this mechanism points out that the short-time growth is characterized by a general Hamiltonian structure of the system and not by the regular to chaotic regimes observed in the studied spin chains.
\par
We consider Pauli operator in transverse direction of the coupling and $\mathcal{\hat U}_x=\hat U_{xx}\hat U_z$ where, $\hat U_{xx}=\exp\big[-i \tau (J_x \hat H_{xx}+h_{x}\hat H_{x})\big]$ and $ \hat U_{z}=\exp( -i \tau h_{z} \hat H_{z})$.  Using Eq.~\ref{HBC}, the Heisenberg evolution of the spin operator $\hat \sigma_z^l$ is obtained:
\begin{eqnarray}
\hat \sigma_z^l(n)&=&(\hat U_z^{\dagger} \hat U_{xx}^{\dagger})^n\hat \sigma_z^l (\hat U_{xx} \hat U_{z})^n,
\end{eqnarray}
after applying first kick $\hat \sigma_z^l(1)$ is  
\begin{eqnarray}
\hat \sigma_z^l(1)&=& \hat U_z^{\dagger} \hat U_{xx}^{\dagger}\hat \sigma_z^l \hat U_{xx} \hat U_{z}, \nonumber \\ 
&=&\hat U_z^{\dagger}(\hat \sigma_z^l+i\tau[\hat H_{xx}+\hat H_{x}, \hat \sigma_z^l]+\frac{(i \tau)^2}{2!}[\hat H_{xx}+\hat H_{x}, [\hat H_{xx}+\hat H_{x}, \hat \sigma_z^l]+\cdots)\hat U_z, \nonumber \\
&=&\hat \sigma_z^l+ i\tau \Big(\hat U_z^{\dagger} (-2i(\hat \sigma_x^{l-1}\hat \sigma_y^{l} + \hat \sigma_y^l\hat \sigma_x^{l+1}+  \hat \sigma_y^l ) \hat U_{z}\Big)+\cdots, \nonumber \\
&=&\hat \sigma_z^l+2\tau \Big(\hat U_{z}^{\dagger}( \hat \sigma_x^{l-1}\hat \sigma_y^{l}+\hat \sigma_y^l\hat \sigma_x^{l+1}+\hat \sigma_y^l ) \hat U_{z}\Big)+\cdots, \nonumber \\ 
&=& \hat \sigma_z^l+2\tau \Big(\hat \hat \sigma_x^{l-1}\hat \sigma_y^{l}+ \hat \sigma_y^l\hat \sigma_x^{l+1}+\hat \sigma_y^l+i\tau(-2i[\hat \sigma_x^{l-1}\hat \sigma_x^{l}+\hat \sigma_y^{l-1}\hat \sigma_y^{l}+ \hat \sigma^l_x\hat \sigma^{l+1}_x+\hat \sigma^l_y\hat \sigma^{l+1}_y + \hat \sigma_x^l])+\cdots\Big)+\cdots, \nonumber \\ 
&=& \hat \sigma_z^l+ \Big(2\tau(\hat \sigma_x^{l-1}\hat \sigma_y^{l}+ \hat \sigma_y^l\hat \sigma_x^{l+1}+\hat \sigma_y^l)+(2\tau)^2(\hat \sigma_x^{l-1}\hat \sigma_x^{l}+\hat \sigma_y^{l-1}\hat \sigma_y^{l}+ \hat \sigma^l_x\hat \sigma^{l+1}_x+\hat \sigma^l_y\hat \sigma^{l+1}_y + \hat \sigma_x^l)+\cdots\Big)+\cdots .
\end{eqnarray}
We apply second kick then $\hat \sigma_z^l(2)$ will be
 \begin{eqnarray}
\hat \sigma_z^l(2)&=&\hat U_z^{\dagger} \hat U_{xx}^{\dagger}\Big(\hat \sigma_z^l+ \Big(2\tau(\hat \sigma_x^{l-1}\hat \sigma_y^{l}+ \hat \sigma_y^l\hat \sigma_x^{l+1}+\hat \sigma_y^l)+(2\tau)^2(\hat \sigma_x^{l-1}\hat \sigma_x^{l}+\hat \sigma_y^{l-1}\hat \sigma_y^{l}+ \hat \sigma^l_x\hat \sigma^{l+1}_x+\hat \sigma^l_y\hat \sigma^{l+1}_y + \hat \sigma_x^l)+\cdots\Big)+\cdots \Big)\hat U_{xx} \hat U_{z},   \nonumber \\  
&=&\hat \sigma_z^l+ \Big(4\tau( \hat \sigma_y^{l-1}\hat \sigma_x^{l}+\hat \sigma_y^l\hat \sigma_x^{l+1}+\hat \sigma_y^l)+(2\tau)^2(\hat \sigma^{l-1}_x\hat \sigma^{l}_x+\hat \sigma^{l-1}_y\hat \sigma^{l}_y + \hat \sigma^l_x\hat \sigma^{l+1}_x+\hat \sigma^l_y\hat \sigma^{l+1}_y + \hat \sigma_x^l) \nonumber \\
&+&(2\tau)^2(\hat U_z^{\dagger} \hat U_{xx}^{\dagger}(\hat \sigma_y^{l-1}\hat \sigma_y^{l}+\hat \sigma_y^l\hat \sigma_y^{l+1})\hat U_{xx}^{\dagger} \hat U_{z}^{\dagger})+\cdots\Big)+\cdots  \end{eqnarray}
From the above equation, we extract the coefficient of $\tau^2$ which contain $\hat \sigma_y^{l+2}$ term. This is given as
 \begin{eqnarray}
 (2\tau)^2(\hat U_z^{\dagger} \hat U_{xx}^{\dagger}(\hat \sigma_y^{l-1}\hat \sigma_y^{l}+\hat \sigma_y^l\hat \sigma_y^{l+1})\hat U_{xx}^{\dagger} \hat U_{z}^{\dagger})&=&(2\tau)^2\hat U_z^{\dagger}\Big(\hat \sigma_y^l\hat \sigma_y^{l+1}+i\tau[\hat H_{xx}+\hat H_x, \hat \sigma_y^l\hat \sigma_y^{l+1}]+\cdots\Big)\hat U_z, \\ \nonumber
 &=&(2\tau)^2\Big(\cdots -2\tau\hat \sigma_y^{l-1}\hat \sigma_z^{l}\hat \sigma_y^{l+1} -2\tau\hat \sigma_y^l\hat \sigma_z^{l+1}\hat \sigma_y^{l+2}+\cdots \Big).
 \end{eqnarray}
 For $m=l+2$, $C_z^{l,m}(2)=64\tau^6$
We apply the third kick then $\hat \sigma_z^l(3)$ will be
 \begin{eqnarray}
 \hat \sigma_z^l(3)&=&\hat U_z^{\dagger} \hat U_{xx}^{\dagger}\Big(\hat \sigma_z^l+ \Big(4\tau(\hat \sigma_y^{l-1}\hat \sigma_x^{l}+ \hat \sigma_y^l\hat \sigma_x^{l+1}+\hat \sigma_y^l)+(2\tau)^2( \hat \sigma^{l-1}_x\hat \sigma^{l}_x+\hat \sigma^{l-1}_y\hat \sigma^{l}_y +\hat \sigma^l_x\hat \sigma^{l+1}_x+\hat \sigma^l_y\hat \sigma^{l+1}_y + \hat \sigma_x^l)+\cdots\Big)+\cdots\Big)\hat U_{xx} \hat U_{z}, \nonumber  \\  \nonumber
&=&\Big[\hat \sigma_z^l+ \Big(6\tau( \hat \sigma_y^{l-1}\hat \sigma_x^{l}+\hat \sigma_y^l\hat \sigma_x^{l+1}+\hat \sigma_y^l)+(2\tau)^2(\hat \sigma^{l-1}_x\hat \sigma^{l}_x+\hat \sigma^{l-1}_y\hat \sigma^{l}_y + \hat \sigma^l_x\hat \sigma^{l+1}_x+\hat \sigma^l_y\hat \sigma^{l+1}_y + \hat \sigma_x^l)+\cdots\Big)+\cdots \Big].
\end{eqnarray}
For $\Delta l=1$, $m=l+1$ dominating exponent of the power-law of the OTOC will be 
\begin{eqnarray}
C_z^{l,l+1}(1)&=&4\tau^2\langle \phi_0\vert[( \hat \sigma_y^l\hat \sigma_x^{l+1}+\hat \sigma_y^l), \hat \sigma_z^m]^2\vert\phi_0\rangle, \\ \nonumber &=&4\langle\phi_0\vert(-i\hat \sigma_y^l\hat \sigma_y^{l+1})^2\vert\phi_0\rangle =4 \tau^2.
\end{eqnarray}
Similarly, $C_z^{l,l+1}(2)=16\tau^2$ and $C_z^{l,l+1}(3)=36\tau^2$.

For $\Delta l=2$, $m=l+2$, dominating exponent of the power-law of the OTOC will be 
$C_z^{l,l+2}(1)=0$, $C_z^{l,l+2}(2)=64\tau^6$.
For $\Delta l=2$, $m=l+2$.
This power-law growth approximately matches the dynamic region of the  Eq.~(\ref{C_tmotoc_nint}).  
 \section{Time evolution of LMOTOC}
 \label{HBC_LMOTOC}
We consider Pauli operator in longitudinal direction of the coupling. Using Eq.~\ref{HBC}, the Heisenberg evolution of the spin operator $\hat \sigma_x^l$ is obtained:
\begin{eqnarray}
\hat \sigma_x^l(n)&=&(\hat U_z^{\dagger} \hat U_{xx}^{\dagger})^n\hat \sigma_x^l (\hat U_{xx} \hat U_{z})^n, \nonumber
\end{eqnarray}
after applying first kick $\hat \sigma_x^l(1)$ is  \begin{eqnarray}
\hat \sigma_x^l(1)&=& \hat U_z^{\dagger} \hat U_{xx}^{\dagger}\hat \sigma_x^l \hat U_{xx} \hat U_{z}, \nonumber \\
&=&\hat U_z^{\dagger}(\hat \sigma_x^l+i\tau[\hat H_{xx}+\hat H_{x}, \hat \sigma_x^l]+\frac{(i \tau)^2}{2!}[\hat H_{xx}+\hat H_{x}, [\hat H_{xx}+\hat H_{x}, \hat \sigma_x^l]+\cdots)\hat U_z. \nonumber
\end{eqnarray}
 Since, $[\hat H_{xx}+\hat H_{x}, \hat \sigma_x^l]=0$, then 
\begin{eqnarray}
\hat \sigma_x^l(1)&=& \hat U_z^{\dagger} \hat \sigma_x^l  \hat U_{z}, \nonumber \\
&=&\hat \sigma_x^l+i\tau[\hat H_{z}, \hat \sigma_x^l]+\frac{(i \tau)^2}{2!}[\hat H_{z}, [\hat H_{z}, \hat \sigma_x^l]+\cdots=
\hat \sigma_x^l(1)=\hat \sigma_x^l- 2\tau \hat \sigma_y^l+\cdots.
\end{eqnarray}
We apply second kick then $\hat \sigma_x^l(2)$ will be
\begin{eqnarray}
\hat \sigma_x^l(2)&=&\hat U_z^{\dagger} \hat U_{xx}^{\dagger}\hat \sigma_x^l(1) \hat U_{xx} \hat U_{z}, \nonumber \\  
&=&U_z^{\dagger} \hat U_{xx}^{\dagger}(\hat \sigma_x^l- 2\tau \hat \sigma_y^l+\cdots)\hat U_{xx} \hat U_{z},\nonumber \\  
&=&\Big(U_z^{\dagger} \hat U_{xx}^{\dagger}\hat \sigma_x^l \hat U_{xx} \hat U_{z} - 2\tau  U_z^{\dagger} \hat U_{xx}^{\dagger}\hat \sigma_y^l\hat U_{xx} \hat U_{z}+\cdots\Big), \nonumber \\
&=&\Big(\hat \sigma_x^l- 2\tau \hat \sigma_y^l-2\tau  U_z^{\dagger} (\hat \sigma_y^l-2\tau (\hat \sigma_y^{l-1} \hat \sigma_z^{l} +\hat \sigma_z^l \hat \sigma_y^{l+1}+  \hat \sigma_z^l)+\cdots) \hat U_{z}+\cdots \Big), \nonumber\\ 
&=&\Big(\hat \sigma_x^l- 2\tau \hat \sigma_y^l-2\tau  [\hat \sigma_y^l-2\tau \hat \sigma_y^{l-1} \hat \sigma_z^{l}-2\tau \hat \sigma_z^l\hat \sigma_y^{l+1}-2\tau  \hat \sigma_z^l+i\tau(-2i \hat \sigma_x^l-2\tau  (-2i)\hat \sigma_z^l \hat \sigma_y^{l+1})+\cdots] +\cdots \Big),\nonumber \\ 
&=&\Big(\hat \sigma_x^l- 4\tau \hat \sigma_y^l+(2\tau)^2 (\hat \sigma_z^l\hat \sigma_y^{l+1}+\hat \sigma_x^l+  \hat \sigma_z^l)+(2\tau )^3 \hat \sigma_z^l \hat \sigma_y^{l+1}) +\cdots \Big). 
\end{eqnarray}
We apply third kick then $\hat \sigma_x^l(3)$ will be
\begin{eqnarray}
\hat \sigma_x^l(3)&=&\hat U_z^{\dagger} \hat U_{xx}^{\dagger}\Big(\hat \sigma_x^l- 4\tau \hat \sigma_y^l+(2\tau)^2 (\hat \sigma_z^l\hat \sigma_y^{l+1}+ \hat \sigma_x^l+ \hat \sigma_z^l)+(2\tau)^3 \hat \sigma_z^l \hat \sigma_y^{l+1}) +\cdots \Big) \hat U_{xx} \hat U_{z}, \nonumber \\ 
&=&\Big(\hat \sigma_x^l- 6\tau \hat \sigma_y^l+2(2\tau )^2 (\hat \sigma_z^l\hat \sigma_y^{l+1}+\hat \sigma_x^l+ \hat \sigma_z^l)+2(2\tau )^3 \hat \sigma_z^l \hat \sigma_y^{l+1}) +\cdots \Big)+\cdots
\end{eqnarray}
Consider $\Delta l=1,$ $m=l+1$
$C_x^{l,l+1}(1)=0$, $C_x^{l,l+1}(2)=64 \tau^6 $, and $C_x^{l,l+1}(3)=256 \tau^6$. Exponent of the power-law approximately matches the  Eq.~(\ref{bformula_lmotoc}). 
\end{widetext}
\end{document}